\documentclass[%
aip,%
amsmath,amssymb,floatfix,
reprint,%
onecolumn,
jcp,%
longbibliography,
]{revtex4-2} 

\usepackage{graphicx}% Include figure files
\usepackage{dcolumn}% Align table columns on decimal point
\usepackage{bm} % bold math % test
\usepackage[usenames,dvipsnames]{xcolor}
\usepackage{soul}

\usepackage[left= 2.9cm, right = 2.9cm, top= 2cm, bottom = 2cm]{geometry}  

\usepackage[english]{babel} 

\usepackage{amsmath,amsthm,latexsym,amssymb,amsfonts,epsfig}
\usepackage{mathtools}
\usepackage{mathrsfs}

\usepackage{xcolor}
\definecolor{OliveGreen}{rgb}{0,0.6,0}

\usepackage[colorlinks=true,citecolor=OliveGreen,linkcolor=blue,urlcolor=blue]{hyperref}

\usepackage{epstopdf}
\usepackage{csquotes}
\usepackage{setspace}
\usepackage{multirow}
\usepackage{gensymb} % For \degres

\allowdisplaybreaks

\usepackage{makecell}
\usepackage{physics}
\usepackage{siunitx}
\usepackage{floatrow}
\usepackage{microtype}
\usepackage{comment}

\emergencystretch=\maxdimen
\begin{document}

% \preprint{APS/123-QED}

\title{Hydrodynamics of an odd active surfer in a chiral fluid \vspace{0.5cm}}

\author{Yuto \textls{Hosaka}}\email{yuto.hosaka@ds.mpg.de}
\affiliation{Max Planck Institute for Dynamics and Self-Organization (MPI-DS), Am Fassberg 17, 37077 G\"{o}ttingen, Germany}

\author{Ramin \textls{Golestanian}} 
\affiliation{Max Planck Institute for Dynamics and Self-Organization (MPI-DS), Am Fassberg 17, 37077 G\"{o}ttingen, Germany}
\affiliation{Institute for the Dynamics of Complex Systems, University of G\"{o}ttingen, 37077 G\"{o}ttingen, Germany}
\affiliation{Rudolf Peierls Centre for Theoretical Physics, University of Oxford, Oxford OX1 3PU, UK}

\author{Abdallah \textls{Daddi-Moussa-Ider}} 
\affiliation{Max Planck Institute for Dynamics and Self-Organization (MPI-DS), Am Fassberg 17, 37077 G\"{o}ttingen, Germany}

\date{July 24, 2023}

\begin{abstract}
We theoretically and computationally study the low-Reynolds-number hydrodynamics of a linear active microswimmer surfing on a compressible thin fluid layer characterized by an odd viscosity.
Since the underlying three-dimensional fluid is assumed to be very thin compared to any lateral size of the fluid layer, the model is effectively two-dimensional.
In the limit of small odd viscosity compared to the even viscosities of the fluid layer, we obtain analytical expressions for the self-induced flow field, which includes non-reciprocal components due to the odd viscosity.
On this basis, we fully analyze the behavior of a single linear swimmer, finding that it follows a circular path, the radius of which is, to leading order, inversely proportional to the magnitude of the odd viscosity.
In addition, we show that a pair of swimmers exhibits a wealth of two-body dynamics that depends on the initial relative orientation angles as well as on the propulsion mechanism adopted by each swimmer.
In particular, the pusher-pusher and pusher-puller-type swimmer pairs exhibit a generic spiral motion, while the puller-puller pair is found to either co-rotate in the steady state along a circular trajectory or exhibit a more complex chaotic behavior resulting from the interplay between hydrodynamic and steric interactions.
Our theoretical predictions may pave the way toward a better understanding of active transport in active chiral fluids with odd viscosity, and may find potential applications in the quantitative microrheological characterization of odd-viscous fluids.
\end{abstract}

\maketitle

%%%%%%%%%%%%%%%%%
\section{Introduction}
%%%%%%%%%%%%%%%%%

The emerging field of soft active matter has recently attracted significant attention among scientists in the bioengineering and biophysics communities~\cite{marchetti2013hydrodynamics, gompper2020}.
Autonomous self-propelling active agents are entities that are capable of converting the energy extracted from their host fluid environment into persistent directed motion and mechanical work.
Over the last few decades, there have been mounting interest and tremendous research efforts in designing and fabricating artificial self-propelled active agents as they set forth as model systems for the physical understanding of out-of-equilibrium phenomena occurring in cellular biology and biochemistry.
Engineered man-made active microswimmers hold great promise for future biomedical and biotechnological applications such as targeted drug delivery~\cite{park2017multifunctional, sharan2021microfluidics, sridhar2022light} and directed transport of curative substances and antibiotic drugs to compromised cancer cells or inflammation sites~\cite{zhong2020photosynthetic, wang2022intelligent}.

At very small scales of motion, viscous effects dominate inertial effects~\cite{happel2012low} so that the behavior of active microswimmers can well be described by Stokes hydrodynamics~\cite{elgeti2015physics, lauga2016bacterial, bechinger2016active}. 
Under these circumstances, microswimmers have to deploy effective swimming strategies to break the time-reversal symmetry imposed by the Stokes flow~\cite{purcell1977}.
A large class of microswimmers are capable of swimming motility by undergoing a non-reciprocal deformation sequence of their shape during a swimming stroke as seen, for instance, in undulatory or amoeboid swimmers~\cite{gilpin2020multiscale, farutin2013amoeboid}.
Meanwhile, phoretic microswimmers achieve autonomous intrinsic self-propulsion through an effective slip profile resulting from local self-generated gradients induced by chemical reactions continuously occurring within a thin interaction layer on their surface~\cite{anderson1989, golestanian2005propulsion, golestanian2007designing,tailleur2022active}.

To better understand the behavior of microswimmers in their complex natural habitat, various model swimmers have been proposed previously.
Notable examples of model microswimmers that have been studied thoroughly in the literature include the classical three-sphere model~\cite{najafi2004,golestanian2008, nasouri2019efficiency, daddi2020tuning} and the squirmer model that self-propels forward by generating tangential surface waves~\cite{lighthill1952, pedley2016squirmers, ito2019swimming}.
Based on these models, the effect of the fluid-mediated hydrodynamic interactions on the clustering mechanism of microswimmer suspension has been investigated~\cite{pooley2007, matas2014hydrodynamic, qi2022emergence, zhang2021effective}.
Low-Reynolds-number hydrodynamics of microswimmers in various geometrical environments has been analyzed, e.g., in the presence of confining walls~\cite{li2014hydrodynamic, lintuvuori2016hydrodynamic, daddi2018swimming}, microfluidic channels~\cite{wioland2016directed, daddi2018state}, viscous drops~\cite{sprenger2020towards, kree2022mobilities}, or deformable interfaces and biological membranes~\cite{nambiar2022hydrodynamics, daddi2019frequency,ota2018three}.
The behavior of non-axially symmetric active swimmers has also been studied in detail, and it has been found that shape anisotropy leads to persistent curvature of trajectories, often resulting in circular swimming~\cite{kummel2013circular, lowen2016, chepizhko2019ideal, chepizhko2022resonant}.

Despite intensive research efforts and various contributions made to unravel the fascinating physics of microswimmers in complex fluid environments, the swimming behavior in chiral environments is only partially understood.
The time-reversal and parity symmetries of chiral fluids are broken as a result of rotating constituents, giving rise to an intrinsic chiral nature of the system under investigation.
In biological systems, for instance, rotating membrane inclusions~\cite{oppenheimer2019rotating}, beating cilia~\cite{han2018spontaneous} and rotating bacterial flagella on active carpets~\cite{uchida2010synchronization, guzman2021active}, \enquote{dancing} chiral algae~\cite{drescher2009dancing}, or spinning starfish embryos~\cite{tan2022odd} can drive the suspending fluid out of equilibrium and support the emergence of chiral structures.
In synthetic systems, particulate suspensions of rotating colloids~\cite{bililign2022motile, han2021,soni2019, debets2023glassy}, ferromagnetic spinners~\cite{kokot2017active}, or inertial vibrots~\cite{scholz2021surfactants} lead to a chiral behavior of the overall host fluid medium.
In particular, if the mean distance between the active constituents is much smaller than the typical spatial scale under consideration, these chiral systems can be regarded as active continuous media that can well be characterized by an \textit{odd viscosity}~\cite{avron1998,banerjee2017}.
In chiral fluids, the typical distance between rotating entities can span over several orders of magnitude, ranging from the nanometer to micrometer scale~\cite{banerjee2017}.

Chiral fluids show several peculiar features such as dissipationless transport~\cite{reichhardt2022, hosaka2021_2}, non-reciprocal response~\cite{hosaka2021,lou2022,lier2022,yuan2023stokesian,hosaka2023pair}, and topological edge flows~\cite{souslov2019,soni2019}.
Odd viscosity is a rheological property that describes chiral fluids~\cite{fruchart2023odd,hosaka2022} and represents a transport coefficient that does not contribute to dissipation of mechanical energy of the fluid~\cite{khain2022}.
The effect of odd viscosity on various types of flows has been examined extensively in the context of microparticle transport~\cite{ganeshan2017,hosaka2021,hosaka2021_2,lier2022} or Langevin dynamics in fluctuating active systems~\cite{yasuda2022}.
Although swimming in fluids with odd viscosity has been partially studied using the geometric theory of low-Reynolds-number locomotion of nearly circular microswimmers~\cite{lapa2014}, to the best of our knowledge, the individual and pair dynamics of microswimmers in an odd fluid have not been addressed so far.
Therefore, detailed investigations in this direction are needed.

In this paper, we theoretically and computationally study the hydrodynamics of active linear swimmers in a compressible fluid layer with odd viscosity, supported by a fluid film underneath.
We analytically obtain the real-space Green's function of the system with non-reciprocal components in the limit of small odd viscosity compared to even viscosities.
Here, we make use of the derived Green's function to obtain the self-generated flow field induced by a model dipolar microswimmer moving in an odd-viscous fluid.
We find that a single linear microswimmer follows a circular trajectory while two nearby microswimmers show nontrivial chiral two-body dynamics that depend on initial relative angles, swimmer types, and magnitude of the relevant odd viscosity coefficients.

The remainder of the paper is organized as follows.
In Sec.~\ref{sec:model}, we introduce a model dipolar microswimmer that propels itself through a fluid via two oppositely directed active point forces acting on both sides of the swimmer.
In Sec.~\ref{sec:mobility}, we recall the previously derived Green's function for the two-dimensional (2D) fluid with odd viscosity and make use of these in Sec.~\ref{sec:single} to determine expressions for the translational and angular velocities of a single swimmer.
Thereupon, we obtain in Sec.~\ref{sec:two} the equations governing dynamics of two hydrodynamically interacting swimmers and fully analyze the pair dynamics in Sec.~\ref{sec:result}.
Finally, Sec.~\ref{sec:summary} contains a summary of the paper, offers further discussions, and outlines some future directions.

\begin{figure}[tb]
\centering
\includegraphics[scale=0.75]{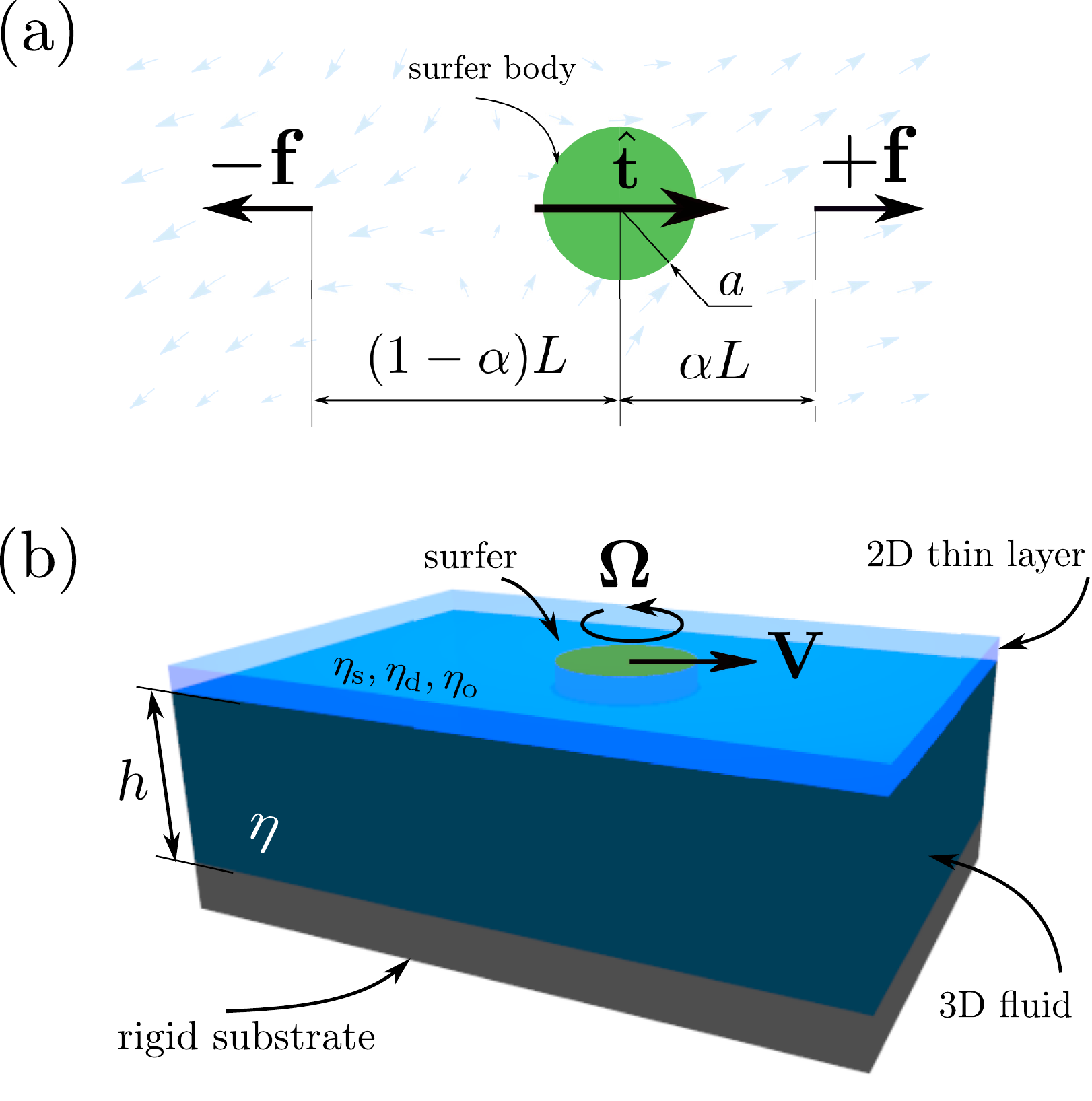} % 0.65
\caption{
(Color online)
(a) 
Schematic illustration of the model microswimmer used in the present study.
The active microswimmer is modeled as a circular disk of radius~$a$ propelled through two active point forces $+\mathbf{f}$ and $-\mathbf{f}$ acting at two point centers separated by a distance~$L$ along the swimmer axis. 
The latter has the unit orientation vector $\hat{\mathbf{t}}$ pointing toward the nearer force center.
Due to the action of the two asymmetrically positioned and oppositely directed point forces, the swimmer can move laterally within the plane of the figure.
(b)
Schematic sketch of an infinitely extended 2D thin layer
of a compressible fluid with 2D shear $\eta_{\rm s}$, dilatational $\eta_{\rm d}$, and odd $\eta_{\rm o}$ viscosities.
The fluid layer overlays a 3D bulk fluid medium of 3D shear viscosity~$\eta$ and film thickness~$h$.
The active microswimmer can move laterally within the 2D fluid layer with translational $\mathbf{V}$ and rotational $\boldsymbol{\Omega}$ velocities.
In the sketch, the thickness of the 3D fluid $h$ appears to be large for illustrative purposes only.
In the actual calculations, we assume a very small film thickness $h$, relative to the overall lateral dimensions of the system, so that despite the 3D representation in the sketch, the fluid effectively forms a thin layer.
\label{fig:model}
}
\end{figure}

%%%%%%%%%%%%%%%%%%%%%%%%%
\section{Model microswimmer}
%%%%%%%%%%%%%%%%%%%%%%%%%
\label{sec:model}

We consider a simple model microswimmer that is composed of a circular disk-like rigid body of radius~$a$ subjected to hydrodynamic drag force.
As a model for the impetus that propels the swimmer forward, we introduce two oppositely directed point forces $\mathbf{f}$ and $-\mathbf{f}$ along the symmetry axis of the swimmer, as schematically illustrated in Fig.~\ref{fig:model}(a).
The behavior of a pusher-type swimmer (extensile) is described by directing the two antagonist point forces away from the disk-like body, while a puller-type swimmer (contractile) can be modeled by directing the two point forces inward.
The model was originally introduced by \textls{Menzel} \textit{et~al.}~\cite{menzel2016dynamical} and has since been used extensively in the context of the statistical description of microswimmer suspensions using dynamical density functional theory approaches~\cite{hoell2017,pessot2018}, or to study the alignment behavior of microswimmers in a nematic liquid crystalline solution~\cite{daddi2018}.
We note that an analogous model that rigorously accounts for the finite-sized shape of the swimmer has further been proposed~\cite{adhyapak2017flow}.

Denoting by $\hat{\mathbf{t}}$ the unit vector describing the orientation of the swimmer and by $\overline{\mathbf{r}}$ the position of the center of the circular body, the right and left point forces acting along the swimming axis are located at positions
$\mathbf{r}_+=\overline{\mathbf{r}} + \alpha L \, \hat{\mathbf{t}}$~~and~~$\mathbf{r}_-=\overline{\mathbf{r}} - (1-\alpha) L \, \hat{\mathbf{t}}$, respectively.
Here, $L$ is the distance between the two point centers, representing a typical size of the swimmer, and $\alpha$ is a dimensionless number chosen to take values between $0$ and $1/2$.
For $\alpha=1/2$ the swimmer is known as a \enquote{shaker}~\cite{vskultety2020swimming}, the situation in which the swimmer pumps the fluid but its position remains unchanged in the absence of hydrodynamic interactions with nearby objects.
In the special limit of $L\to0$, the model for $\alpha=1/2$ reduces to the standard hydrodynamic force dipole commonly used to describe the leading-order behavior of force-free microswimmers~\cite{berke2008hydrodynamic} and enzymatic molecules~\cite{mikhailov2015,agudo2018enhanced,manikantan2020,hosaka2023pair} in fluids.

The fluid velocity at position $\mathbf{r}$ of the flow induced by the two oppositely directed point forces is then obtained via linear superposition as
\begin{align}
\mathbf{v}(\mathbf{r}) = 
\left[ \mathbf{G}(\mathbf{r}-\mathbf{r}_+) - \mathbf{G}(\mathbf{r}-\mathbf{r}_-) \right]\cdot\mathbf{f} \, ,
\label{eq:inducedv}
\end{align}
where $\mathbf{G}(\mathbf{r}-\mathbf{r}_\pm)$ is the Green's function or the second-rank mobility tensor that provides the induced velocity field at position $\mathbf{r}$ due to the point force singularity $\pm \mathbf{f}$ acting on the fluid layer at position $\mathbf{r}_\pm$.
The expression of $\mathbf{G}$ will later be provided in Sec.~\ref{sec:mobility}.
Here, $\mathbf{f} = f\hat{\mathbf{t}}$, such that $f>0$ for a pusher and $f<0$ for a puller.
Restricting ourselves to the situation in which $a \ll L$, the self-induced translational and rotational velocities of the swimmer are obtained to leading order as~\cite{lauga2020,manikantan2020}
\begin{equation}
\mathbf{V} = \left. \mathbf{v}(\mathbf{r}) \right|_{\mathbf{r} = \overline{\mathbf{r}}} \, , \qquad
\boldsymbol{\Omega} = \left. \frac{1}{2} \, \nabla\times \mathbf{v}(\mathbf{r}) \right|_{\mathbf{r} = \overline{\mathbf{r}}}  \, .
\label{eq:vomega}
\end{equation}
Here we have assumed the lowest order of the 2D Fax\'{e}n laws~\cite{oppenheimer2009}, which do not consider the radius $a$ of the swimming body. 
For higher order contributions, the use of Green's function becomes necessary to account for the near field hydrodynamic effect~\cite{pessot2018}.
When the interacting circular swimmers are identical in size, their finite sizes can be included via the Rotne-Prager-Yamakawa approximation, which considers the next-higher order term, i.e., the contribution proportional to $(a/L)^2$~~\cite{rotne1969variational, yamakawa1970transport}.
For non-identical particles, numerical methods are available to calculate hydrodynamic interactions~\cite{de1977hydrodynamic}.
However, in this paper, we do not consider the effect of finite-sized swimming bodies due to the complexity of the mobility tensor and leave such an extension as a topic for future work.
Note that our numerical simulations for a pair of microswimmers include a short-range soft repulsion between the swimmers, which corresponds to the effective size of a swimmer.
This will be explained in more detail later.

%%%%%%%%%%%%%%%%%%%%%%%%%%%%%
\section{2D fluid with odd viscosity}
%%%%%%%%%%%%%%%%%%%%%%%%%%%%%
\label{sec:mobility}

In our fluid model, an infinitely extended 2D thin layer of a \emph{compressible} fluid overlays a 3D bulk fluid medium of a 3D shear viscosity~$\eta$ and film thickness~$h$~\cite{hosaka2021}, as schematically depicted in Fig.~\ref{fig:model}(b).
The underlying bulk fluid is bounded from below by an impermeable planar no-slip wall.
We assume that the 2D fluid layer is in contact with air.
We denote by~$\eta_{\rm d}$ and~$\eta_{\rm s}$ the 2D dilatational and shear viscosities, respectively.
In addition, the thin compressible fluid layer possesses a 2D odd viscosity $\eta_{\rm o}$.
Although we do not focus specifically on the origin of the odd viscosity in this study, it can be attributed, for example, to the spinning motion of self-rotating particles adhering to the thin layer, which violate both time-reversal and parity symmetries~\cite{banerjee2017}.
Assuming that such active rotors are homogeneously dispersed in the 2D fluid layer and their concentration is sufficiently low, the 2D layer can be viewed as a layer of a continuum active chiral fluid with an odd viscosity that is spatially constant.
Thus, the microswimmers moving in the layer, which we will consider later, effectively experience the chiral nature of self-spinning particles through the odd viscosity.
Besides these rotors, microswimmers can also give rise to odd viscosity due to their circular motion~\cite{reichhardt2019active}, but we do not consider such an effect in this study.
Due to its dissipationless nature, $\eta_{\rm o}$ can be either positive or negative, while the regular even viscosities ($\eta_{\rm d}$ and $\eta_{\rm s}$) should always remain positive according to the requirement that the dissipated
energy is positive.
At the microscopic scale, the sign of the odd viscosity can be related to the direction of rotation of the active rotating constituents~\cite{markovich2021}.

There are two main reasons that motivated us to consider a two-layer fluid, consisting of a 2D compressible fluid of odd viscosity in contact with an underlying 3D fluid, as shown in Fig.~\ref{fig:model}(b).
First, in the case of a 2D incompressible fluid, the odd viscosity is absorbed into the pressure term of the Stokes equation~\cite{ganeshan2017}.
This feature makes it impossible to observe the effect of odd viscosity on the hydrodynamic interaction between objects in 2D fluids in the incompressibility regime~\cite{khain2022}.
To solve this problem, one can consider a compressible fluid instead, and such a system can be realized, e.g., by a dilute Gibbs monolayer, consisting of soluble amphiphiles capable of dissolving in the underlying 3D fluid~\cite{barentin1999}.
Second, in a purely 2D fluid without any momentum decay, there is no hydrodynamic screening length in the system, preventing the derivation of a Green's function valid for arbitrary length scales.
For example, by introducing an outer 3D fluid, the screening length is determined by the viscosities of both the 2D and 3D fluids, allowing the full Green's function to be obtained analytically.

Using the lubrication approximation for the 3D fluid, which holds in the limit of very small thickness~$h$ compared to any lateral size of the system, the momentum balance equation for the 2D fluid can be written as~\cite{hosaka2021}
\begin{subequations} \label{eq:Stokescomp}
\begin{align}
\eta_{\rm s}\nabla^2\mathbf{v} +\eta_{\rm d}\nabla (\nabla\cdot\mathbf{v}) + \eta_{\rm o}\nabla^2 (\boldsymbol{\epsilon}\cdot\mathbf{v})
 -\frac{h}{2}\nabla p
-\frac{\eta}{h}\mathbf{v}
+\mathbf{F} =\mathbf{0}&,\label{eq:Stokes}\\
\nabla\cdot\mathbf{v} = \frac{h^2}{6\eta}\nabla^2p&,
\label{eq:comp}
\end{align}
\end{subequations}
where $p$ is the 3D pressure, $\boldsymbol{\epsilon}$ stands for the 2D Levi-Civita (permutation) tensor such that $\epsilon_{xx}=\epsilon_{yy}=0$ and $\epsilon_{xy}=-\epsilon_{yx}=1$, and $\mathbf{F}$ is any force density acting on the 2D fluid.
Note that the gradient of the 2D pressure is absent in Eq.~(\ref{eq:Stokes}) because we have assumed that the 2D fluid layer quickly equilibrates with the 3D one, and hence the 2D pressure is homogeneous in space, as considered in Refs.~\cite{barentin1999, hosaka2021}.
The 4th and 5th terms in Eq.~(\ref{eq:Stokes}) represent the force exerted by the 3D fluid on the 2D layer, and Eq.~(\ref{eq:comp}) relates the divergence of the in-plane velocity to the Laplacian of the 3D pressure (see Ref.~\cite{barentin1999} for more details on the derivation of these expressions).
The hydrodynamic inverse screening lengths, beyond which the thin fluid layer exchanges momentum with the underlying bulk fluid medium, are defined as
\begin{equation}
\kappa = \sqrt{\frac{\eta}{h \eta_{\mathrm{s}}}} \, , \qquad 
\lambda = \sqrt{\frac{2 \eta}{h \bar{\eta}}} \, ,
\label{kappa_lambda}
\end{equation}
where $\bar{\eta} = \left( \eta_{\mathrm{s}}+\eta_{\mathrm{d}} \right)/2$ denotes an average even viscosity.

The linear hydrodynamic response of the 2D fluid layer is described by the Green's function $\mathbf{G}(\mathbf{r})$ that relates the applied force to the induced velocity:
\begin{align}
    \mathbf{v}(\mathbf{r}) = \int {\rm d}^2\mathbf{r}^\prime\, \mathbf{G}(\mathbf{r}-\mathbf{r}^\prime)\cdot \mathbf{F}(\mathbf{r}^\prime).
\end{align}
To derive $\mathbf{G}(\mathbf{r})$, we solve the hydrodynamic equations~(\ref{eq:Stokescomp}) in Fourier space and obtain $\widetilde{\mathbf{G}} [\mathbf{k}]$, where $\mathbf{k}=(k_x,k_y)$ and the square brackets denote a function in Fourier space.
The Green's function in 2D Fourier space~\cite{bracewell1986fourier} have previously been obtained by \textls{Hosaka} \textit{et~al.}~\cite{hosaka2021}~as
\begin{equation}
\widetilde{\mathbf{G}} [\mathbf{k}] 
= \frac{\eta_{\mathrm{s}}\left(k^{2}+\kappa^{2}\right) \mathbf{k}_\| \, \mathbf{k}_\| + 2\bar{\eta}\left(k^{2}+\lambda^{2}\right) \mathbf{k}_\perp \,  \mathbf{k}_\perp-\eta_{\mathrm{o}} k^{2} \boldsymbol{\epsilon} }{2\eta_{\mathrm{s}}\bar{\eta}\left(k^{2}+\kappa^{2}\right)\left(k^{2}+\lambda^{2}\right)+\eta_{\mathrm{o}}^{2} k^{4}},
\label{eq:G}
\end{equation}
where we have defined two orthogonal unit vectors $\mathbf{k}_\|=\left(k_{x} / k, k_{y} / k\right)$ and $\mathbf{k}_\perp=\left(-k_{y} / k, k_{x} / k\right)$.
We refer the reader to Ref.~\cite{hosaka2021} for the full derivation.
In the limit of $\eta_{\rm d}\to\infty$ [and thus $\lambda\to0$; c.f.\ Eq.~\eqref{kappa_lambda}], the Green's function reduces to that of an incompressible fluid of the shear viscosity $\eta_{\rm s}$~\cite{ramachandran2011,oppenheimer2010}, given by
$\widetilde{\mathbf{G}} [\mathbf{k}] =
\mathbf{k}_\perp \,  \mathbf{k}_\perp / \left( \eta_{\rm s}(k^2+\kappa^2) \right)$, for which the contribution due to the odd viscosity vanishes.

Upon 2D inverse Fourier transformation of Eq.~(\ref{eq:G}), the Green's function can conveniently be expressed in real space expression as~\cite{doi1988}
\begin{equation}
\mathbf{G}(\mathbf{r})=C_{1}(r) \, \mathbf{I} + C_{2}(r) \, \hat{\mathbf{r}}\hat{\mathbf{r}} + C_{3}(r) \, \boldsymbol{\epsilon} \, ,
\label{eq:gmobility}
\end{equation}
where $\mathbf{I}$ is the second-rank identity tensor, $\hat{\mathbf{r}}=\mathbf{r}/r$ is a unit vector, and the three radial functions $C_i(r)$, $i \in \{1,2,3\}$, are given in the form of infinite integrals over the wavenumber~$k$; see Ref.~\cite{hosaka2021} for their full expressions.
To simplify our theoretical developments and to be able to make analytical progress, we assume for simplicity in the following that $\eta_{\rm d} = 3\eta_{\rm s}$ so that $\kappa = \lambda$.
Accordingly, the three radial functions take the forms
\begin{subequations} \label{eq:c}
\begin{align} 
C_{1}(r)&= \frac{1}{2 \pi \eta_{\rm s}} \int_{0}^{\infty} {\rm d} k \, \frac{k\left(k^{2}+\kappa^{2}\right)}{4\left(k^{2}+\kappa^{2}\right)^{2}+\mu^{2} k^{4}} \left( 4 J_{0} (kr)-\frac{3}{kr} \, J_{1} (kr) \right) \, , 
\label{eq:c1}\\
 C_{2}(r)&= \frac{3}{2 \pi \eta_{\rm s}} \int_{0}^{\infty} {\rm d} k \, \frac{ k\left(k^{2}+\kappa^{2}\right)}{4\left(k^{2}+\kappa^{2}\right)^{2}+\mu^{2} k^{4}} \left( \frac{2}{kr}\, J_{1} (kr) - J_{0}(kr) \right) \, , 
 \label{eq:c2}\\
 C_{3}(r)&= -\frac{\mu}{2 \pi \eta_{\rm s}} \int_{0}^{\infty} {\rm d} k \, \frac{k^{3} J_{0} (kr)}{4\left(k^{2}+\kappa^{2}\right)^{2}+\mu^{2} k^{4}} \, ,
 \label{eq:c3}
\end{align}
\end{subequations}
with $J_n(\cdot)$ denoting Bessel function of the first kind of order~$n$~\cite{abramowitz1964handbook}.
Here, we have defined for convenience the ratio
\begin{equation}
    \mu = \frac{\eta_{\rm o}}{\eta_{\rm s}} \, .
\end{equation}
It is worth noting that the Green's function stated by Eq.~\eqref{eq:gmobility} will involve a contribution stemming from the non-reciprocal term proportional to~$\boldsymbol{\epsilon}$ if and only if $\mu \ne 0$.
See Appendix~\ref{app:Ci} for the expressions of $C_1$, $C_2$, and $C_3$ when $\eta_{\rm d} \neq 3\eta_{\rm s}$.

\begin{figure}
\centering
\includegraphics[scale=0.5]{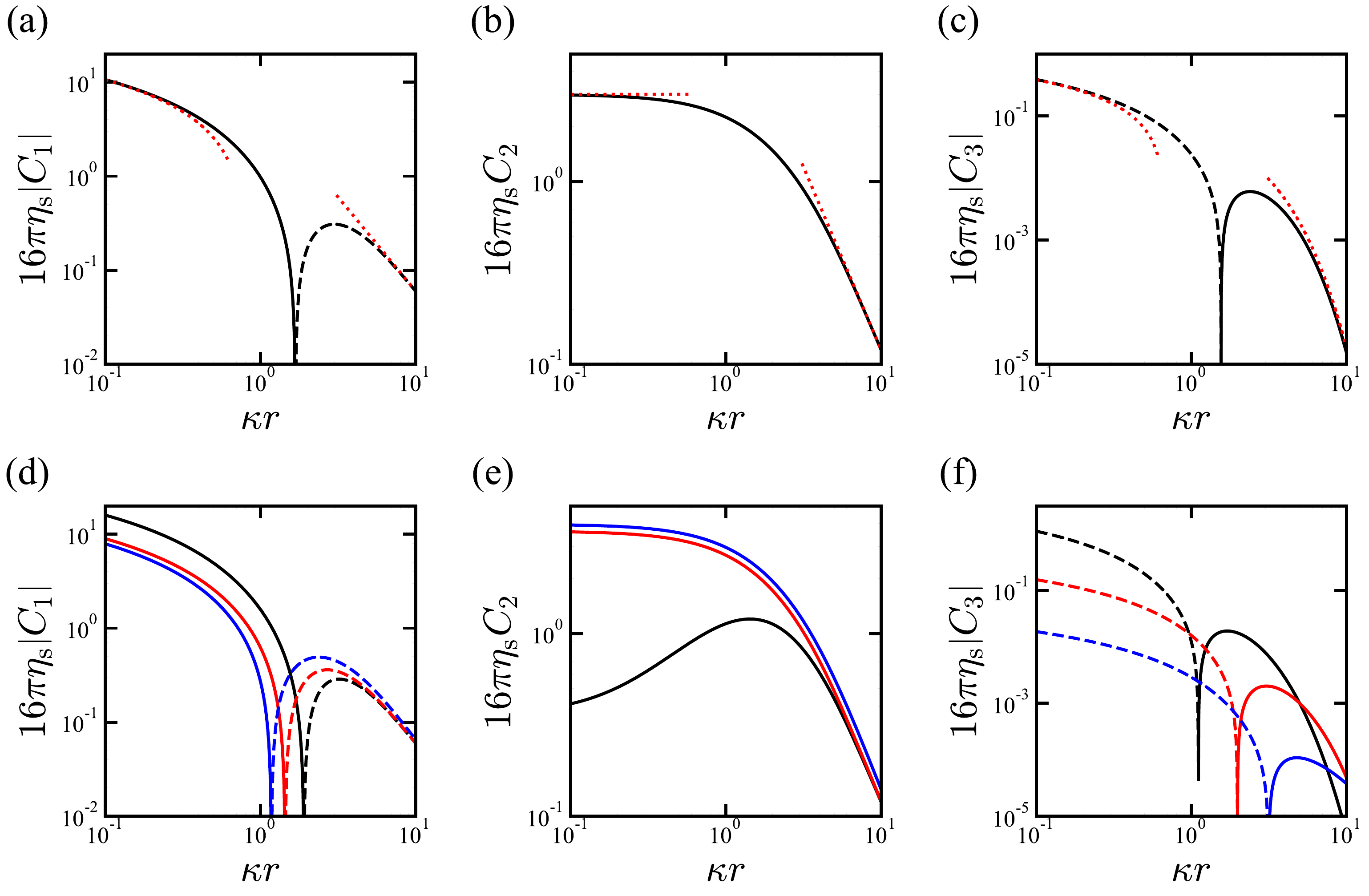}
\caption{
(Color online)
Log-log plots of the scaled radial functions defining the Green's function given by Eq.~\eqref{eq:gmobility}.
The radial functions of (a)~$C_1$, (b)~$C_2$, and (c)~$C_3$ when $\eta_{\rm d} = 3\eta_{\rm s}$ [see Eqs.~\eqref{eq:cmull1} for their expressions] as a function of $\kappa r$, while keeping the ratio of the odd to shear viscosity $\mu=0.1$.
The asymptotic behaviors of these functions, as analytically
given by Eqs.~\eqref{eq:c123small} and (\ref{eq:c123large}) in the near and far fields, respectively, are plotted with red dotted lines.
The radial functions of (d)~$C_1$, (e)~$C_2$, and (f)~$C_3$ for $\eta_{\rm d}/\eta_{\rm s}=0.1,10,$ and $100$ (black, red, and blue lines) [see Eqs.~\eqref{eq:appc1c2c3} for their expressions] as a function of $\kappa r$, while keeping $\mu=0.1$.
Since $C_1$ and $C_3$ take negative values in the limits of large and small $\kappa r$ (shown by the dashed lines), respectively, in the log-log plots, we have plotted their absolute values instead in (a), (c), (d), and (f).}
\label{fig:ci}
\end{figure}

In the limit of $\mu \ll 1$, the infinite integrals given by Eqs.~\eqref{eq:c} can be evaluated analytically,
\begin{subequations} \label{eq:cmull1}
\begin{align} 
C_{1}(r)&\simeq \frac{1}{2 \pi \eta_{\rm s}}
\left[K_0(\kappa r) - \frac{3}{4(\kappa r)^2}+ \frac{3K_1(\kappa r)}{4\kappa r}\right]
\label{eq:c1mull1}
, \\
 C_{2}(r)&\simeq \frac{3}{4 \pi \eta_{\rm s}}
\left[ \frac{1}{(\kappa r)^2} - \frac{K_2(\kappa r)}{2}\right]
\label{eq:c2mull1}
, \\
 C_{3}(r)&\simeq \frac{\mu}{16\pi \eta_{\rm s}}
\left[
 \kappa r K_1(\kappa r) -2K_0(\kappa r)
 \right], 
 \label{eq:c3mull1}
\end{align}
\end{subequations}
wherein $K_n(\cdot)$ is the modified Bessel function of the second kind of degree~$n$~\cite{abramowitz1964handbook}.
In the present contribution, we assume that the overall displacement of the microswimmer is much smaller than the hydrodynamic screening lengths such that $\kappa r\ll1$.
Then the asymptotic expressions of Eqs.~\eqref{eq:cmull1} in this limiting situation are expressed by
\begin{subequations} \label{eq:c123small}
\begin{align} 
C_{1}(r)&\simeq \frac{1}{32 \pi \eta_{\rm s} }
\left[ 10 \ln \left( \frac{2}{\kappa r}\right) -3-10\gamma \right]
, 
 \\
 C_{2} &\simeq \frac{3}{16 \pi \eta_{\rm s}} 
 \, , \\
 C_{3}(r)&\simeq-\frac{\mu}{16 \pi \eta_{\rm s}}
\left[
2 \ln\left(\frac{2}{\kappa r}\right)
-1-2\gamma
 \right],
 \end{align}
\end{subequations}
where $\gamma = 0.5772157 \dots$ is Euler's constant (also sometimes called the Euler-Mascheroni constant).
In the opposite limit of $\kappa r\gg1$, the asymptotic expressions of Eqs.~\eqref{eq:cmull1} are obtained as
\begin{equation} 
C_{1}(r) \simeq -\frac{3h}{8 \pi \eta r^2} \, , 
 \qquad
 C_{2}(r) \simeq \frac{3h}{4 \pi \eta r^2} \, ,  
 \qquad
 C_{3}(r) \simeq \mu \, \sqrt{\frac{\pi\kappa r}{2}} \, e^{-\kappa r} \, , 
 \label{eq:c123large}
\end{equation}
where $C_1$ and $C_2$ both show an algebraic dependence on~$r$ with a power of $-2$, whereas $C_3$ decays exponentially upon increasing $\kappa r$.

In Fig.~\ref{fig:ci}(a)-(c), we plot the variations of the scaled radial functions $C_1$, $C_2$, and $C_3$ as given by Eqs.~\eqref{eq:cmull1} as a function of $\kappa r$.
Since $C_1$ and $C_3$ take negative values in the limits of large and small $\kappa r$, respectively, we have plotted the corresponding absolute values as dashed lines.
The asymptotic expressions in the near and far fields, as respectively given by Eqs.~\eqref{eq:c123small} and (\ref{eq:c123large}), are also shown as red dotted lines, which provide good approximations especially for $\kappa r\ll1$ and $\kappa r\gg1$.
At small length scales, all the three radial functions depend only weakly on~$\kappa r$, while at large scales, both $C_1$ and $C_2$ show an algebraic decay as $\left(\kappa r\right)^{-2}$ and $C_3$ shows a sharp exponential decay.

While our focus in this study is on the effect of odd viscosity on swimming dynamics, here we briefly demonstrate how the Green's function depends on variations in the even viscosities, namely, in the case where $\eta_{\rm d}\neq 3\eta_{\rm s}$.
In Fig.~\ref{fig:ci}(d)-(f), we plot $C_1, C_2,$ and $C_3$ as a function of $\kappa r$ for various values of $\eta_{\rm d}/\eta_{\rm s}$, while keeping $\mu=0.1$ as before.
Since from Eq.~(\ref{kappa_lambda}), the two screening lengths are related by $\lambda^{-1}=(\kappa^{-1}/2)\sqrt{1+\eta_{\rm d}/\eta_{\rm s}}$, we vary $\eta_{\rm d}/\eta_{\rm s}$ to see the effect of the even viscosities.
As shown in Fig.~\ref{fig:ci}(d) and (f), the crossover positions between the positive and negative values of $C_1$ and $C_3$ shift as $\eta_{\rm d}/\eta_{\rm s}$ is varied because the screening lengths are dependent of the even viscosities.
Figure~\ref{fig:ci}(e) shows that $C_2$ exhibits nonmonotonic behavior, with a maximum value at $\kappa r\approx2$, for $\eta_{\rm d}/\eta_{\rm s}=0.1$; for larger values of $\eta_{\rm d}/\eta_{\rm s}$, it shows a monotonic decreasing dependence, similar to the case $\eta_{\rm d}/\eta_{\rm s}=3$ in Fig.~\ref{fig:ci}(b).
Since the flow field does not depend on the odd viscosity in the incompressibility regime $(\eta_{\rm d}\to\infty)$, $C_3$ decreases extensively for $\eta_{\rm d}/\eta_{\rm s}=100$ compared to the case $\eta_{\rm d}/\eta_{\rm s}=0.1$, as in Fig.~\ref{fig:ci}(f).

%%%%%%%%%%%%%%%%%%%%%%%%%%%%%%%%%%%
\section{single swimmer dynamics}
%%%%%%%%%%%%%%%%%%%%%%%%%%%%%%%%%%%
\label{sec:single}

\begin{figure}
\floatbox[{\capbeside\thisfloatsetup{capbesideposition={right,center},capbesidewidth=5cm}}]{figure}[\FBwidth]
{\caption{(Color online) Quiver plot of the 2D velocity field $\mathbf{v}(x,y)$ generated by a pusher-type microswimmer ($f>0$) positioned at the origin of coordinates and directed along the horizontal axis such that $\hat{\mathbf{t}}~=~\hat{\mathbf{e}}_x$.
Here, we set a viscosity ratio $\mu=\eta_{\rm o}/\eta_{\rm s}=1$ and consider a swimmer of size $\kappa L=0.1$ and an asymmetry parameter $\alpha=2/5$.
The black arrows denote point forces that act at positions $\mathbf{r}_+=(0.4L,0)$ and $\mathbf{r}_-=(-0.6L,0)$.
The colorbar shows the magnitude of the self-induced flow velocity scaled by $f/(16\pi\eta_{\rm s})$.
The swimmer body of circular shape is not shown here for clarity.
The flow field induced by a puller for the same geometric and physical parameters can be obtained by inverting the directions of the velocity vectors. 
}
\label{fig:flow}}
{\includegraphics[scale=0.95]{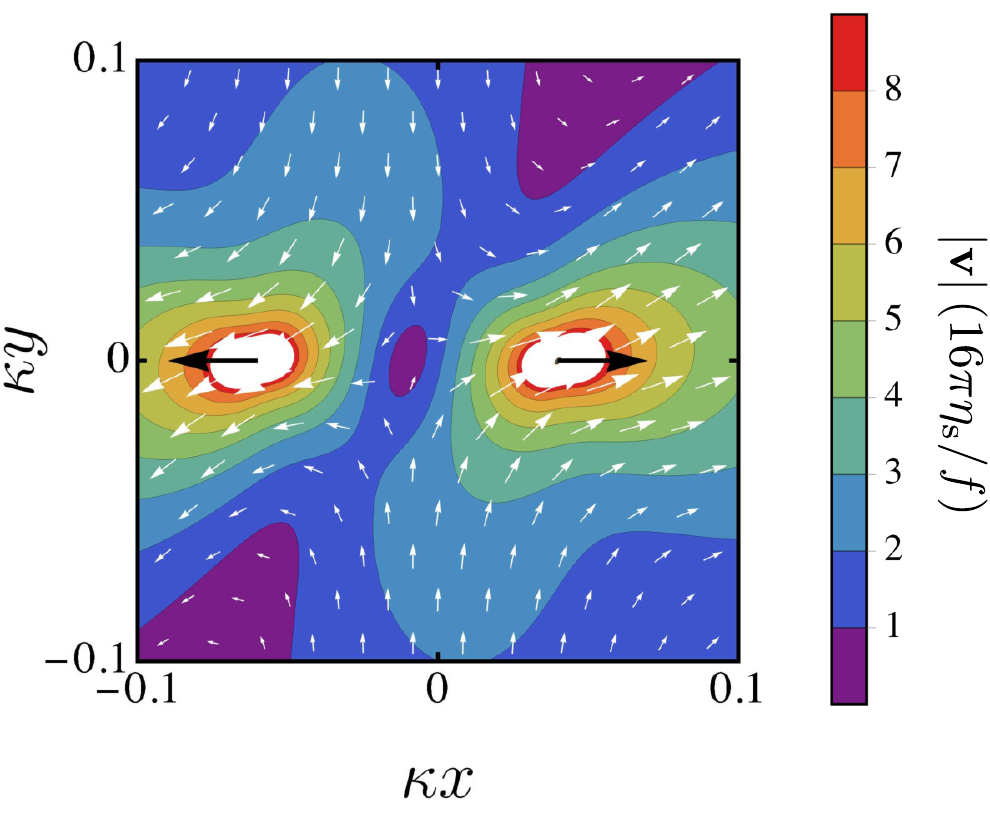}}
\end{figure}

Having introduced a relevant model microswimmer and summarized in a nutshell the solution for the Green's function in a 2D fluid with odd viscosity, we next obtain expressions of the self-induced velocity field of the swimmer.
We then proceed to deriving approximate expressions of the translational and angular velocities of a single swimmer in the limiting cases of $\mu \ll 1$ and $\kappa r = \lambda r \ll 1$, and discuss the resulting dynamics of a single swimmer in an odd-viscous fluid.

%%%%%%%%%%%%%%%%%%%%%%%%%%%%
\subsection{Velocity field}

Substituting Eqs.~\eqref{eq:c} into Eq.~(\ref{eq:inducedv}), we can evaluate numerically the velocity field induced by the model swimmer for an arbitrary values of $\mu=\eta_{\rm o}/\eta_{\rm s}$.
In Fig.~\ref{fig:flow}, we present a quiver plot and contour diagram of the flow field induced by a pusher $(f>0)$.
The swimmer is centered at the origin and is directed along the horizontal axis with a unit orientation vector $\hat{\mathbf{t}} = \hat{\mathbf{e}}_x$.
Here, we consider a microswimmer of typical size $\kappa L=0.1$ and asymmetry parameter $\alpha=2/5$.
To increase the effect of odd viscosity on the overall flow field, we choose a relatively large ratio of odd to shear viscosity and set $\mu=1$.

Equation~(\ref{eq:gmobility}) implies that the non-reciprocal term proportional to~$\boldsymbol{\epsilon}$ would cause fluid motion not only along the lateral direction but also along the direction transverse to the orientation vector~$\hat{\mathbf{t}}$.
Accordingly, the vertical component of the fluid flow induced by the two active forces is more pronounced in a fluid with odd viscosity than in a fluid without, as shown in Fig.~\ref{fig:flow}.
Specifically, the flow induced by the force center at~$\mathbf{r}_+$ has a bias along the positive vertical axis, whereas a bias toward the negative vertical axis arises around the force center at~$\mathbf{r}_-$.
Consequently, the resulting self-induced fluid flow shows a shear component that causes an overall anti-clockwise rotation of the swimmer.
An analogous flow behavior is obtained for a puller $(f<0)$ with~$\mu<0$.
In contrast to that, for a pusher with $\mu<0$ or a puller with~$\mu>0$, the swimmer undergoes a clockwise rotation.
For~$\mu=0$, transverse flow lines along the vertical axis vanish and the flow symmetry with respect to the horizontal is recovered.
In addition, the rotation component of the flow in this case vanishes, resulting in a straight swimming trajectory, as can be seen later in the next subsection.

%%%%%%%%%%%%%%%%%%%%%%%%%%%%%%%%%%
\subsection{Swimming velocities}

When $\mu\ll1$, we can obtain the velocity field generated by a model swimmer analytically.
Defining for convenience the abbreviation $R_\pm = \left| \mathbf{R}_\pm \right|= \left| \mathbf{r} - \mathbf{r}_\pm \right|$ to denote the distance from the force centers, the self-generated flow field induced by a microswimmer moving in an odd-viscous fluid can be cast in the form
\begin{equation}
\mathbf{v}(\mathbf{r}) = \frac{5f}{16\pi\eta_{\rm s}}
\left[
\ln\left(\frac{R_-}{R_+}\right)
\left(\mathbf{I}-\frac{2}{5}\mu \boldsymbol{\epsilon} \right)
+
\frac{3}{5} \left( \frac{\mathbf{R}_+\mathbf{R}_+}{R_+^2} - \frac{\mathbf{R}_-\mathbf{R}_-}{R_-^2} \right)
\right]
\cdot\hat{\mathbf{t}} \, .
\label{Eq:VeloField} 
\end{equation}

\begin{figure}
\centering
\includegraphics[scale=0.49]{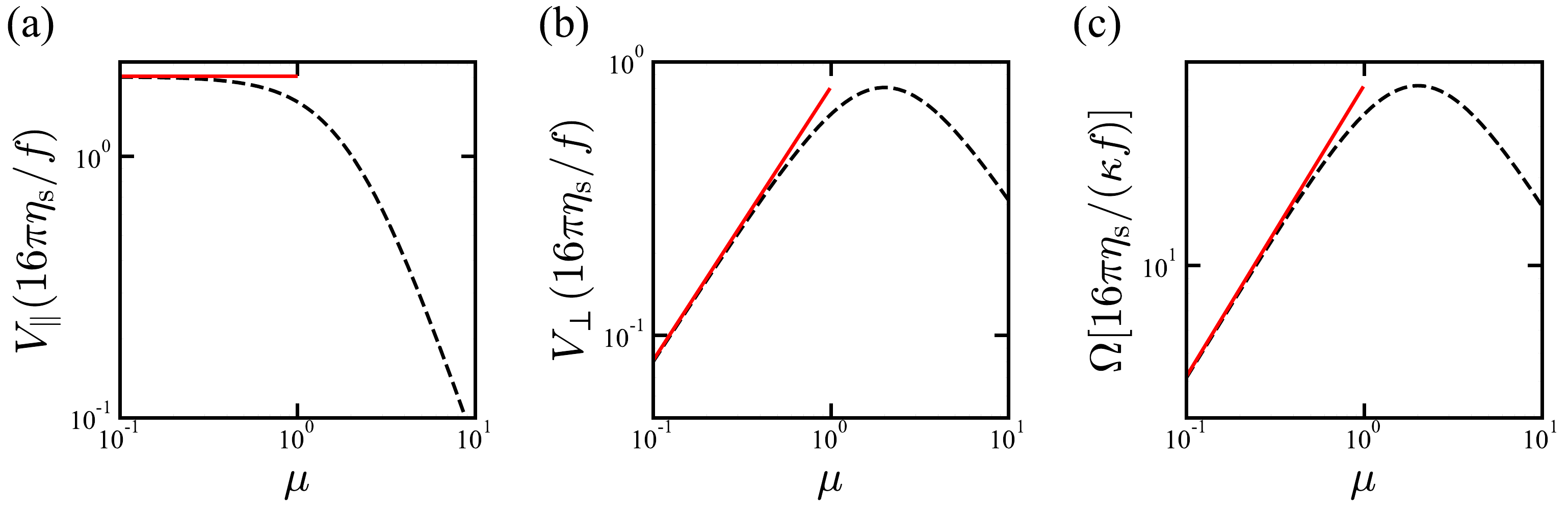}
\caption{
(Color online)
Log-log plots of the scaled self-propulsion speed along the direction (a)~parallel $V_\|=\mathbf{V}\cdot\hat{\mathbf{t}}$ and (b)~perpendicular $V_\perp=\mathbf{V}\cdot\hat{\mathbf{n}}$ to the orientation axis as a function of the viscosity ratio $\mu = \eta_\mathrm{o}/\eta_\mathrm{s}$.
(c)~Corresponding log-log plot of the scaled angular velocity $\Omega=|\mathbf{\boldsymbol{\Omega}}|$ normal to the thin fluid layer.
Translational and rotational velocities are made dimensionless by scaling by $f/(16\pi\eta_{\rm s})$ and $f\kappa/(16\pi\eta_{\rm s})$, respectively.
Analytical expressions as given by Eq.~\eqref{eq:speed-rotation-rate} are shown as red solid lines, while black dashed lines represent the full expressions of the swimming velocity and rotation rate obtained by numerical integration of Eqs.~\eqref{eq:c}.
Here, we set $\kappa L = 0.1$ and choose an asymmetry parameter of the swimmer $\alpha=2/5$.
\label{fig:vxvyo}
}
\end{figure}

We now set $\hat{\mathbf{t}} = \left(\cos\theta,\sin\theta\right)$ where $\theta$ is defined as the angle between the $x$-axis and the orientation vector of the swimmer.
In addition, we define the unit vector $\hat{\mathbf{n}} = \left( -\sin\theta, \cos\theta \right)$ normal to the swimming axis.
It follows from Eq.~(\ref{eq:vomega}) that the translational swimming speed and rotation rate of the swimmer are obtained to leading order in $a/L$ as
\begin{equation}
\mathbf{V} = 
\frac{5\beta f}{16\pi\eta_{\rm s}} 
\left( \hat{\mathbf{t}} + \frac{2}{5}\mu \hat{\mathbf{n}} \right) , \qquad
\boldsymbol{\Omega} = 
\frac{\mu f}{16\pi\eta_{\rm s}\alpha(1-\alpha)L} \, \hat{\mathbf{e}}_z ,
\label{eq:speed-rotation-rate}
\end{equation}
where we have defined
\begin{equation}
    \beta = \ln\left( \frac{1}{\alpha} - 1 \right) ,
\end{equation}
which only take positive values.
Here, $\hat{\mathbf{e}}_z$ is a unit vector directed along the $z$-direction normal to the flat thin layer.
For a non-vanishing odd viscosity, the swimmer undergoes additionally a drift motion along the direction normal to the orientation axis.
This drift is accompanied by a spinning motion that is clockwise for $f\eta_\mathrm{o} <0$ and anti-clockwise for $f\eta_\mathrm{o} >0$.
Notably, the non-reciprocal term proportional to $\boldsymbol{\epsilon}$ in Eq.~(\ref{eq:gmobility}) leads to a non-vanishing angular velocity even when $\alpha=1/2$.
As the swimmer changes from asymmetric $(\alpha\to0)$ to symmetric $(\alpha=1/2)$ shapes, the rotation rate $\Omega = \left| \boldsymbol{\Omega} \right|$ monotonically decreases to reach a minimum value of magnitude $\left|\mu f\right|/ \left(4\pi\eta_\mathrm{s} L\right)$.

Figure~\ref{fig:vxvyo} shows the variation of the components of the translational and rotational velocities upon varying the viscosity ratio~$\mu$, while keeping $\alpha=2/5$ and $\kappa L=0.1$. % as before.
We observe that $V_\parallel=\mathbf{V}\cdot\hat{\mathbf{t}}$ [Fig.~\ref{fig:vxvyo}(a)] is to leading order independent of~$\mu$, as can be inferred from Eq.~(\ref{eq:speed-rotation-rate}), and decreases monotonically as~$\mu$ increases.
In contrast to that, both $V_\perp=\mathbf{V}\cdot\hat{\mathbf{n}}$ and~$\Omega$ [Fig.~\ref{fig:vxvyo}(b) and (c)] increase linearly for $\mu \ll 1$ to reach a maximum value before they eventually decrease as $\mu$ increases.
For $\mu\gg1$, $V_\perp$ and~$\Omega$ show an algebraic dependence on~$\mu$ with a power of $-1$, which can readily be deduced from the scaling behavior of $\mu$ in Eqs.~\eqref{eq:c}.
This non-monotonic behavior suggests that the maximum speed of the swimmer is reached when the odd viscosity relative value is of the same order of magnitude with the shear viscosity.

%%%%%%%%%%%%%%%%%%%%%%%%%%%%%%%
\subsection{Equation of motion}

We denote by $\varphi(t)$ the angle between the translational velocity $\mathbf{V} = V_\parallel \hat{\mathbf{t}} + V_\perp \hat{\mathbf{n}}$ and the $x$-axis.
The equations of motion for the swimmer position $\mathbf{r}(t)=(x(t),y(t))$ can be expressed as~\cite{lauga2020}
\begin{equation}
\dot{\mathbf{r}} (t) = V\hat{\mathbf{p}}(\varphi(t)) \, ,\qquad
\dot{\varphi} (t) = \Omega \, ,
\label{eq:eom2}
\end{equation} 
with dot standing for a temporal derivative.
Moreover, $V = \left| \mathbf{V} \right|$ and $\hat{\mathbf{p}}(\varphi(t))=(\cos\varphi(t),\sin\varphi(t))$ is the unit vector pointing along the swimmer velocity.
Integrating the equation for the rotational degree of freedom, we obtain $\varphi(t) = \varphi_0+\Omega t$.
It can be shown that $\cos\theta = \operatorname{sgn} \left(f\right) \left[ \cos\varphi + (2 \mu/5) \sin\varphi \right]$ so that $\theta$ differs from $\varphi$ in the general case of non-vanishing~$\mu$ for the pusher case.
Nonetheless, we note that $\dot{\theta} = \dot{\varphi}$.

By using the complex number notation for the sake of convenience and setting $z= x+iy$, Eq.~(\ref{eq:eom2}) for the translational degree of freedom can then be written as
$\dot{z}=Ve^{i(\varphi_0+\Omega t)}$,
the solution of which is given by 
\begin{equation}
z(t) = z_0 - i \, \frac{V}{\Omega} \, e^{i\varphi_0}(e^{i\Omega t}-1) \, ,
\end{equation}
where $z_0 = x_0 + iy_0$.
Using Euler's relation, explicit expressions for swimmer's displacement are obtained as
\begin{equation}
x(t) = x_0 + \frac{V}{\Omega} \left[\sin\varphi(t)-\sin\varphi_0 \right] ,\qquad
y(t) = y_0 - \frac{V}{\Omega} \left[\cos\varphi(t)-\cos\varphi_0 \right].
\label{eq:eom}
\end{equation}
Eliminating $\varphi(t)$ from Eq.~(\ref{eq:eom}) yields the equation of a circle, namely,
$\left[ x(t) - \bar{x}\right]^2 + \left[y(t) - \bar{y}\right]^2
=R^2$, wherein
\begin{equation}
R = \left|\frac{V}{\Omega}\right|
=\frac{5\beta L}{\left|\mu\right|} \, \alpha(1-\alpha) \, ,
\label{eq:radius}
\end{equation}
is the radius of the circular path and 
$\left( \bar{x}, \bar{y} \right) =
\left( x_0 - \left(V/\Omega\right) \sin\varphi_0 \, , y_0 + \left(V/\Omega\right) \cos\varphi_0 \right)$ is the position of the center of the circle on the $xy$ plane.
Interestingly, the radius reaches a maximum value given by $R_\mathrm{max} = \left(5L/|\mu|\right) \, \alpha^* \left( 1-\alpha^* \right) / \left( 1-2\alpha^* \right)$ such that $\alpha^*$ is the root of the equation $\left(1-2\alpha^*\right) \beta^* - 1 = 0$.
Using numerical evaluations, we obtain $\alpha^*\approx 0.18$ and $R_\mathrm{max} \approx 1.12\, L/\left|\mu\right|$.

%%%%%%%%%%%%%%%%%%%%%%%%%%%%%%%%%

%%%%%%%%%%%%%%%%%%%%%%%%%%%%%%%%%%%
\section{Two-swimmer dynamics}
%%%%%%%%%%%%%%%%%%%%%%%%%%%%%%%%%%%
\label{sec:two}

\begin{figure}[tb]
\centering
\includegraphics[scale=2]{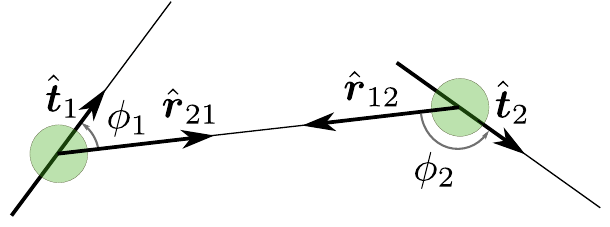} % 1.4
\caption{
(Color online)
Schematic illustration of two microswimmers that interact with each other through far-field hydrodynamic interactions.
We denote by $\hat{\mathbf{t}}_i$ the unit orientation vector of the $i$th swimmer. 
In addition, $\mathbf{r}_{12} = \bar{\mathbf{r}}_1- \bar{\mathbf{r}}_2$ is the vector pointing from swimmer~2 to swimmer~1, $\ell = \left| \mathbf{r}_{12} \right|$ is the distance separating the two swimmers, and $\hat{\mathbf{r}}_{12} = -\hat{\mathbf{r}}_{21} = \mathbf{r}_{12}/\ell$.
Here, $\phi_i$ stands for the angle between $\hat{\mathbf{r}}_{ji}$ and $\hat{\mathbf{t}}_i$.
\label{fig:two}
}
\end{figure}

Having investigated in detail the dynamics of a single swimmer, we next examine the behavior of two hydrodynamically interacting microswimmers in an odd-viscous fluid, as schematically depicted in Fig.~\ref{fig:two}.
The velocity field induced by the $i$th swimmer reads
\begin{equation}
\mathbf{v}_i(\mathbf{r}) = 
[\mathbf{G}(\mathbf{r}-\mathbf{r}_{i}^+) - \mathbf{G}(\mathbf{r}-\mathbf{r}_{i}^-)]\cdot\mathbf{f}_i \, ,
\label{eq:inducedvi}
\end{equation}
for $i \in \{1,2\}$, where $\mathbf{r}_i^+ = \overline{\mathbf{r}}_i + \alpha L \, \hat{\mathbf{t}}_i$ and $\mathbf{r}_i^-= \overline{\mathbf{r}}_i  -(1-\alpha)L \, \hat{\mathbf{t}}_i$ represent for each swimmer the positions of the right and left force centers, respectively.
Here, $\hat{\mathbf{t}}_i=(\cos\theta_i,\sin\theta_i)$ and $\mathbf{f}_i = f_i\hat{\mathbf{t}}_i$, and the vector pointing from the swimmer $j$ to the swimmer $i$ is $\mathbf{r}_{ij} = \overline{\mathbf{r}}_i - \overline{\mathbf{r}}_j$ with $\left|\mathbf{r}_{ij}\right| = \ell$.
The total flow field induced by the two swimmers can then be obtained by linearly superposing the flow fields induced by each swimmer.
Specifically,
\begin{equation}
\mathbf{v} %^{\rm tot}
(\mathbf{r})= \sum_{i=1}^{2} \mathbf{v}_i(\mathbf{r}) \, ,
\label{eq:totalv}
\end{equation}
where the flow velocity field induced by each single swimmer, $\mathbf{v}_i$, is stated by Eq.~\eqref{Eq:VeloField}.

For Stokes flows, velocities and forces are linearly related to each other via the hydrodynamics mobility functions~\cite{kim2013microhydrodynamics}.
Defining $\mathbf{K}_{ij}^+ = \mathbf{r}_{ij} - \alpha L \hat{\mathbf{t}}_j$ and $\mathbf{K}_{ij}^- = \mathbf{r}_{ij} + \left(1-\alpha\right) L \hat{\mathbf{t}}_j$, the translational velocities of the two swimmers are obtained as
\begin{equation}
    \begin{pmatrix}
        \mathbf{V}_1 \\
        \mathbf{V}_2
    \end{pmatrix}
    = 
    \begin{pmatrix}
        \boldsymbol{\Lambda}^\mathrm{S} & \boldsymbol{\Lambda}_{12}^\mathrm{P} \\
        \boldsymbol{\Lambda}_{21}^\mathrm{P}  & \boldsymbol{\Lambda}^\mathrm{S}
    \end{pmatrix}
    \cdot
    \begin{pmatrix}
        \mathbf{f}_1 \\
        \mathbf{f}_2
    \end{pmatrix} ,
    \label{eq:twoV}
\end{equation}  
where the self-interaction part reads
\begin{equation}
    \boldsymbol{\Lambda}^\mathrm{S} = \frac{5\beta}{16\pi\eta_{\rm s}}
\left(\mathbf{I}-\frac{2}{5} \, \mu \boldsymbol{\epsilon} \right) , 
\end{equation}
and the contribution resulting from the pair-interaction part is given by
\begin{equation}
    \boldsymbol{\Lambda}_{ij}^\mathrm{P} =
    \frac{5}{16\pi\eta_{\rm s}}
    \ln \frac{\left| \mathbf{K}_{ij}^- \right|}{\left| \mathbf{K}_{ij}^+ \right|}
    \left(\mathbf{I}-\frac{2}{5} \, \mu \boldsymbol{\epsilon} \right) 
    + \frac{3}{16\pi\eta_\mathrm{s}} \left( \frac{\mathbf{K}_{ij}^+ \mathbf{K}_{ij}^+}{\left| \mathbf{K}_{ij}^+ \right|^2} 
    - \frac{\mathbf{K}_{ij}^- \mathbf{K}_{ij}^-}{\left| \mathbf{K}_{ij}^- \right|^2} \right) .
    \label{eq:pairv}
\end{equation}
The derivation of Eq.~(\ref{eq:twoV}) is shown in Appendix~\ref{app:vomega}.

Similarly, the angular velocities of the pair of swimmers normal to the plane of the thin layer are obtained as
\begin{equation}
    \begin{pmatrix}
        \Omega_1 \\
        \Omega_2
    \end{pmatrix}
    = 
    \begin{pmatrix}
        k^\mathrm{S} & k_{12}^\mathrm{P} \\
        k_{21}^\mathrm{P} & k^\mathrm{S}
    \end{pmatrix}
    \begin{pmatrix}
        f_1 \\
        f_2
    \end{pmatrix} ,
    \label{eq:twoOmega}
\end{equation}
with the self-interaction part
\begin{equation}
    k^\mathrm{S} = \frac{\mu}{16\pi\eta_{\rm s}\alpha(1-\alpha)L} \, , 
\end{equation}
and the pair-interaction part
\begin{align}
k_{ij}^\mathrm{P} &= 
\frac{1}{16\pi\eta_{\rm s}} \left( \frac{\mathbf{K}_{ij}^-}{\left|\mathbf{K}_{ij}^-\right|^2}
-\frac{\mathbf{K}_{ij}^+}{\left|\mathbf{K}_{ij}^+\right|^2} \right)
\cdot
\left(\mu \mathbf{I} + 4\boldsymbol{\epsilon} \right)\cdot\hat{\mathbf{t}}_j \, .
\label{eq:pairomega}
\end{align}
More details about the derivation of Eq.~(\ref{eq:twoOmega}) are given in Appendix~\ref{app:vomega}.

In terms of the polar angle $\phi_i$ between $\mathbf{r}_{ji}$ and $\hat{\mathbf{t}}_i$, such that $\mathbf{r}_{ij} = \ell \left( \cos\phi_j \hat{\mathbf{t}}_j - \sin\phi_j \hat{\mathbf{n}}_j \right)$ with the unit vector $\hat{\mathbf{n}}_i = \left( -\sin\theta_i, \cos\theta_i \right)$, the pair-interaction coupling coefficient can be written as 
\begin{equation}
    k_{ij}^\mathrm{P} = \frac{1}{16\pi\eta_{\rm s}}
\left[
\frac{\ell \left( 4\sin\phi_j + \mu\cos\phi_j \right)+\mu (1-\alpha)L}{\ell^2+2(1-\alpha)L\ell\cos\phi_j + (1-\alpha)^2L^2}
-\frac{\ell \left( 4\sin\phi_j + \mu\cos\phi_j \right)-\alpha \mu L}{\ell^2-2\alpha L\ell\cos\phi_j + \alpha^2L^2}
\right] .
\end{equation}
Since $\mathbf{r}_{21} = -\mathbf{r}_{12}$, it can readily be shown that $\phi_2 - \phi_1 = \theta_2-\theta_1+\pi$.

%%%%%%%%%%%%%%%%%%%%%%%%%%%%%%%%%
\section{Pair dynamics simulations}
%%%%%%%%%%%%%%%%%%%%%%%%%%%%%%%%%
\label{sec:result}

To investigate microswimmer pair dynamics, we perform numerical simulations with different initial angles $\phi_1$ and~$\phi_2$.
For that aim, we solve numerically for $\bar{\mathbf{r}}_i$ and $\phi_i$ the nonlinear dynamical equations of motion (\ref{eq:twoV}) and~(\ref{eq:twoOmega}) that provide the variations of the translational and rotational degrees of freedom, respectively.
In addition to hydrodynamic interactions between the two swimmers, the simulations are carried out by including additionally a short-range soft repulsion, also sometimes called excluded volume interaction, to prevent particle overlap, via~\cite{manikantan2020}
\begin{equation}
\mathbf{U}_{ij} = \frac{2U e^{-\nu\left(\ell-L\right)}}{1+e^{-\nu\left(\ell-L\right)}} \, 
\hat{\mathbf{r}}_{ij} \, ,
\label{eq:soft}
\end{equation}
where $\hat{\mathbf{r}}_{ij} = \mathbf{r}_{ij}/\ell$, and $U$ and~$\nu$ are positive numbers.
Steric interactions are introduced through a contact velocity that decays exponentially with $\nu \left(\ell-L\right)$.
The far-field motion is driven solely by 2D hydrodynamic interactions described by Eqs.~(\ref{eq:pairv}) and (\ref{eq:pairomega}).

In total, we obtain from Eqs.~(\ref{eq:twoV}) and (\ref{eq:soft}) the translational velocities of the swimmer $1$ relative to the swimmer $2$ as
\begin{align}
    \dot{\bar{\mathbf{r}}}_1
    =
    \frac{5}{16\pi\eta_{\rm s}}
    \left[
    \left(\mathbf{I}-\frac{2}{5} \, \mu \boldsymbol{\epsilon} \right)\cdot
    \left(
    \beta\mathbf{f}_1
    +
    \ln \frac{\left| \mathbf{K}_{12}^- \right|}{\left| \mathbf{K}_{12}^+ \right|}
    \mathbf{f}_2
    \right)
    + 
    \frac{3}{5}
    \left( \frac{\mathbf{K}_{12}^+ \mathbf{K}_{12}^+}{\left| \mathbf{K}_{12}^+ \right|^2} 
    - \frac{\mathbf{K}_{12}^- \mathbf{K}_{12}^-}{\left| \mathbf{K}_{12}^- \right|^2} \right)
    \cdot\mathbf{f}_2
    \right]
    +
    \mathbf{U}_{12} \, ,
\end{align}
where we have replaced $\mathbf{V}_1$ with $\dot{\bar{\mathbf{r}}}_1$ and similarly for the translational velocity of the swimmer $2$.
Each swimmer rotates at a rate equal to half the vorticity of the flow velocity induced by itself and the other swimmer, as given by Eq.~(\ref{eq:twoOmega}).
For the dynamics of the relative angles $\phi_1$ and $\phi_2$, there is also an additional effect of the rotation of the connecting vector $\mathbf{r}_{12}$, when the induced velocity has a component that is perpendicular to the vector $\mathbf{r}_{12}$~\cite{manikantan2020, hosaka2023pair}.
In total, the relative angle of the swimmer $1$ to the swimmer $2$ evolves as
\begin{align}
    \dot{\phi}_1
    =
    \frac{\mu f_1}{16\pi\eta_{\rm s}\alpha(1-\alpha)L}
    +
    \frac{1}{16\pi\eta_{\rm s}} \left( \frac{\mathbf{K}_{12}^-}{\left|\mathbf{K}_{12}^-\right|^2}
    -\frac{\mathbf{K}_{12}^+}{\left|\mathbf{K}_{12}^+\right|^2} \right)
    \cdot
    \left(\mu \mathbf{I} + 4\boldsymbol{\epsilon} \right)\cdot\mathbf{f}_2
    +
    \frac{1}{\ell}
    \left(\dot{\bar{\mathbf{r}}}_2-\dot{\bar{\mathbf{r}}}_1\right)
    \cdot\boldsymbol{\epsilon}\cdot\hat{\mathbf{r}}_{21} \, ,
\end{align}
and similarly for $\dot{\phi}_2$.
In the above, the last term represents the relative rotation contribution and it vanishes for the dynamics of the absolute angles $\theta_1$ and $\theta_2$.
Throughout the numerical simulations, we set the asymmetry parameter of the swimmers to $\alpha=2/5$ and consistently choose the viscosity ratio $\mu=0.1$.
We also consider an initial separation distance between the swimmers of $\ell=4L$.

To quantitatively characterize the dynamics of pair swimmers, we introduce a measure of the chirality of the swimming trajectories, using curvature as a metric.
For 2D systems, an instantaneous curvature of a trajectory can be defined as
\begin{align}
    \zeta = \frac{\dot{x}\ddot{y}-\dot{y}\ddot{x}}{(\dot{x}^2+\dot{y}^2)^{3/2}}.
    \label{eq:curvature}
\end{align}
The time-averaged curvature, denoted as $\bar{\zeta}$, serves as a quantitative measure of the chirality of the swimming paths and is used in the following numerical results. 
For a circular path, the chirality takes finite values, whereas for a straight trajectory, it becomes zero.
For instance, the chirality of a single swimmer can be analytically determined from Eq.~(\ref{eq:eom}) as $\bar{\zeta}_0=|\Omega/V|=\left|\mu\right|/[5\beta L \alpha(1-\alpha)]$ that coincides with the inverse of the radius of the circular path in Eq.~(\ref{eq:radius}), as it should.
For the parameters used in the numerical simulations ($\mu=0.1$ and $\alpha=2/5$), one can estimate $L\bar{\zeta}_0\approx0.21$.
For the pair dynamics, the chirality is expected to deviate from this value due to the influence of hydrodynamic interaction from the other swimmer.
It is important to note that the above estimates hold true under the assumption that the trajectories resulting from pair hydrodynamic interactions are essentially deviations from the circular path, as observed for an isolated single swimmer. Our measure of chirality would not be applicable to chiral trajectories if this assumption no longer holds.

\begin{figure}[tb]
\centering
\includegraphics[scale=0.6]{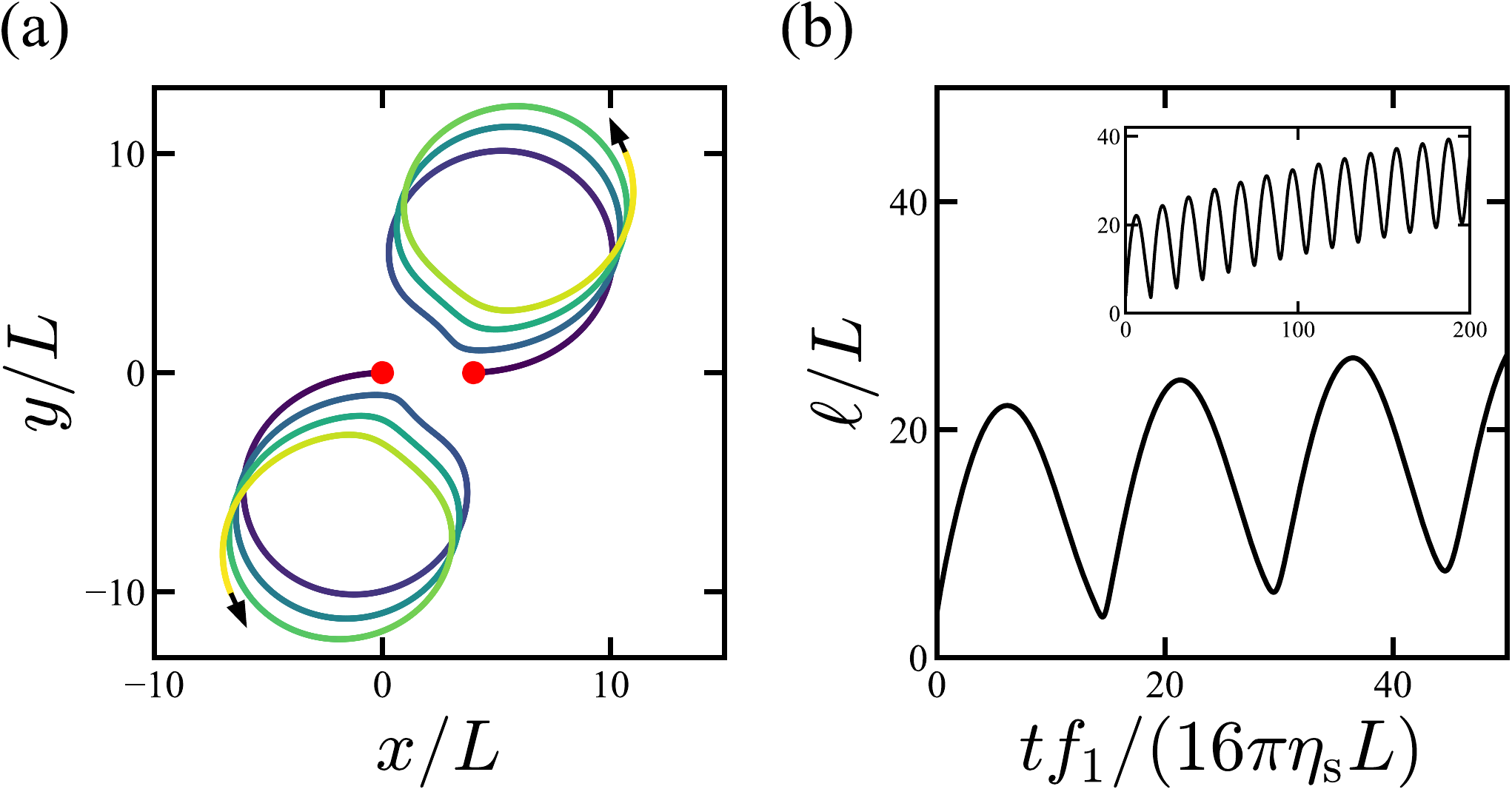} % 1.4
\caption{(Color online) 
Hydrodynamics of a pusher-pusher microswimmer pair moving in an odd-viscous fluid.
(a)~Swimming trajectories describing the positions of the centers of the disks  $\overline{\mathbf{r}}_i$, obtained by numerically integrating the equations of motion governing the dynamics of the two swimmers given by Eqs.~\eqref{eq:twoV} and~\eqref{eq:twoOmega}.
The color in the trajectories gradually changes from blue to yellow over time.
(b)~Corresponding time evolution of the inter-swimmer distance $\ell$.
The swimmer pair displays a diverging spiral evolution during which the mutual distance between the two swimmers increases in time when averaging over an oscillation period. 
Here, we set the ratio of odd to shear viscosity $\mu=0.1$ and an asymmetry parameter of the swimmers $\alpha=2/5$.
The pair is initially located at positions $(x,y)=(0,0)$ and $(4L,0)$ and the initial orientations are set $(\phi_1,\phi_2)=(\pi,\pi)$.
In a panel (b), the inset shows the evolution of $\ell$ over a longer period of time.}
\label{fig:pusherpusher}
\end{figure}

\begin{figure}
\centering
\includegraphics[scale=0.65]{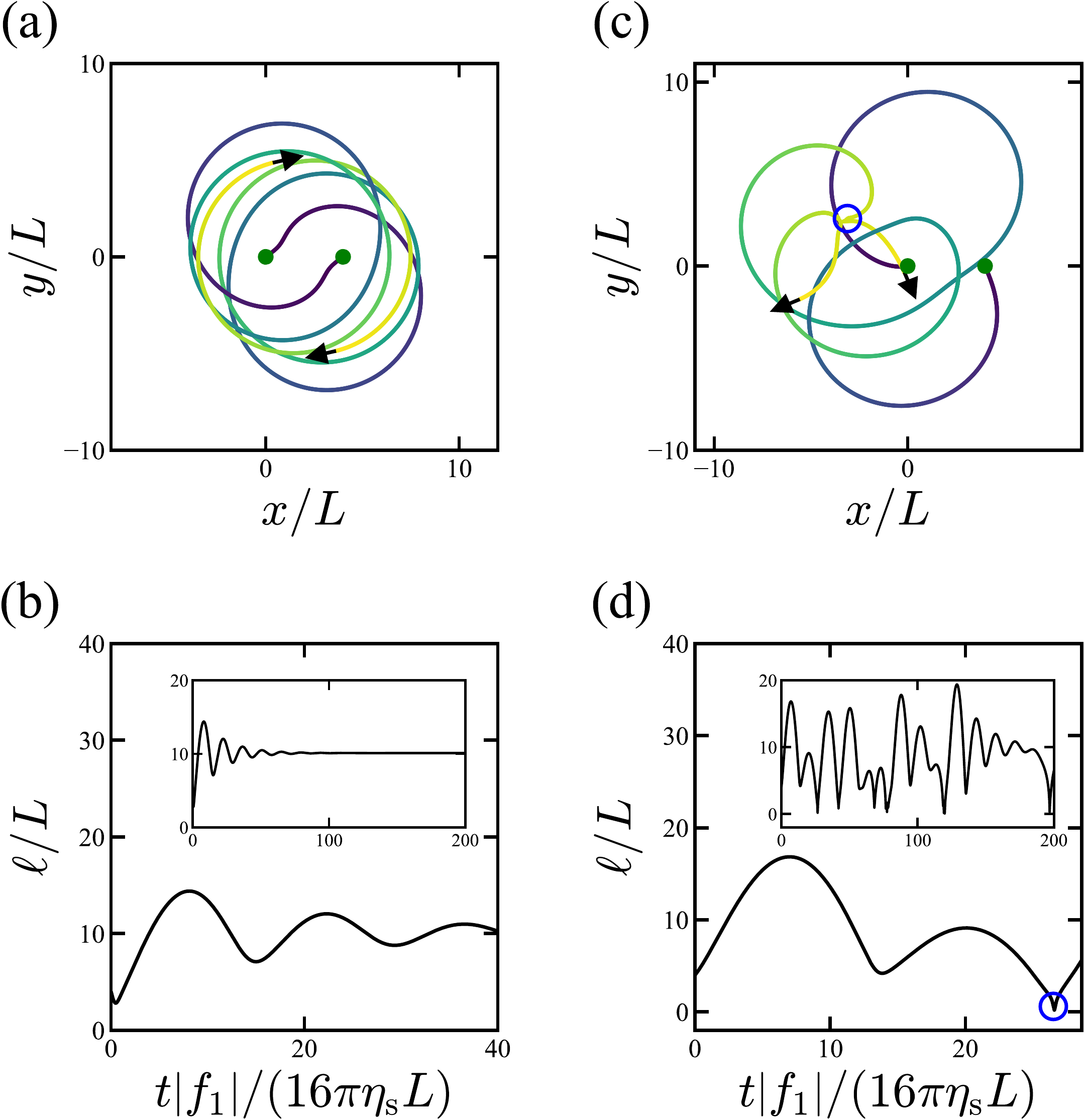} % 0.4
\caption{
Dynamics of a puller-puller swimmer pair initially located at positions $(x,y)=(0,0)$ and $(4L,0)$.
Swimming trajectories described by the position of the swimmers $\overline{\mathbf{r}}_i$ for (a) stable co-orbiting and (c) unstable co-orbiting states.
The color in the trajectories gradually changes from blue to yellow over time.
Corresponding time evolution of the inter-swimmer distance $\ell$ are shown in panels (b) and~(d).
Here, we set $\mu=0.1$ and $\alpha=2/5$.
The initial orientations for (a,b) and (c,d) are set $(\phi_1,\phi_2)=(6\pi/5,6\pi/5)$ and $(2\pi,9\pi/5)$, respectively.
The blue circles in (c,d) correspond to the position and time at which the two swimmers interact sterically.
In panels (b) and (d), the insets show the evolution of $\ell$ over a longer period of time.
}
\label{fig:pullerpuller}
\end{figure}

We find that a pair of self-propelling active swimmers in an odd-viscous fluid shows rich chiral dynamics that depends on the propulsion mechanism of the swimmers as well as on their initial relative orientations $\phi_1$ and $\phi_2$.
The resulting trajectories of the swimmer pairs show typically oscillatory trajectories owing to the chiral nature of the surrounding fluid.
We find that the period-averaged mutual distance between two pushers ($f_1>0$ and $f_2>0$) grows with time while it either shows a chaotic behavior or reaches a stable dynamical state for two pullers ($f_1<0$ and $f_2<0$).
In contrast to that, for a pusher-puller swimmer pair ($f_1 f_2 < 0$), the mutual distance separating two swimmers is generally found to remain constant on average over a period.
A detailed discussion concerning the resulting swimming trajectories and mutual distance between the swimmers will be presented in the following subsections.
The time evolution of the inter-swimmer distance and the swimmers' trajectories are shown in Figs.~\ref{fig:pusherpusher}-\ref{fig:pusherpuller} and in the supplemental video.

%%%%%%%%%%%%%%%%%%%%%%%%%%%%%%%%
\subsection{Pusher-pusher swimmer pair}
\label{sec:puspus}

Figure~\ref{fig:pusherpusher}(a) shows the swimming trajectories of a pusher-pusher microswimmer pair moving on a thin fluid layer characterized by a ratio of odd to shear viscosity of $\mu = 0.1$.
The two swimmers are initially located at positions $(x,y) = (0,0)$ and $(4L,0)$.
The two pushers exhibit rotational motion in an anti-clockwise direction while gradually moving apart.
They follow trajectories that can be described as spirals, or as circular motions that are drifting apart.
We denominate this pattern as a \enquote{diverging spiral} state because the two swimmers are, on average, repel each other.
Correspondingly, the inter-swimmer distance $\ell$ shows non-stationary behavior through an oscillatory dependence on time; see Fig.~\ref{fig:pusherpusher}(b).
The average mutual distance between the two swimmers increases with time and is expected to reach a plateau value when the two swimmers are sufficiently displaced far apart for $\ell \gg L$.
We note that the observed spiral motion is attributed to the non-vanishing non-reciprocal component proportional to~$\boldsymbol{\epsilon}$ in the Green's function, which leads to finite transverse and rotational velocities in the flow fields induced by the two swimmers.

\begin{figure}
\centering
\includegraphics[scale=0.65]{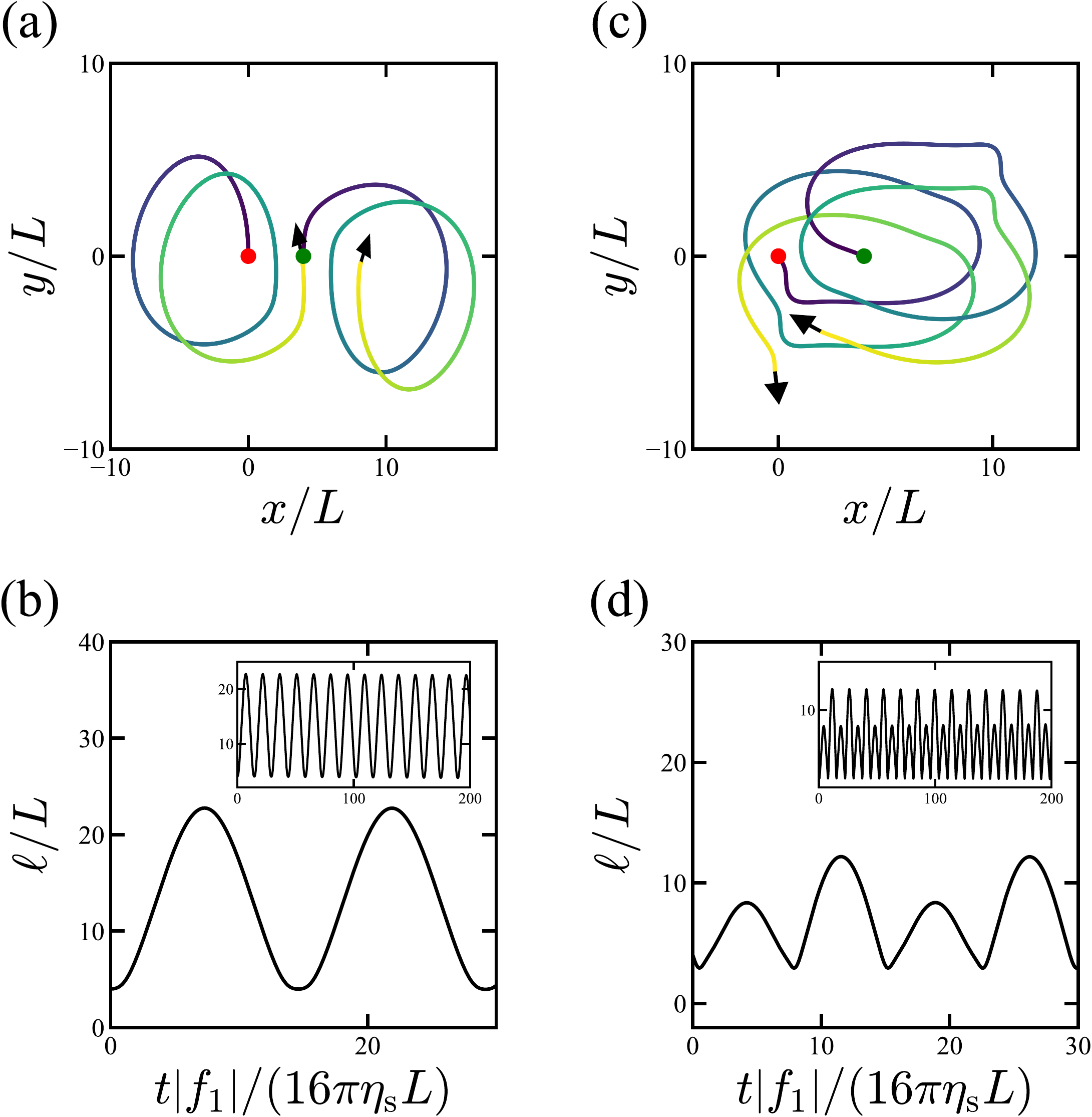}
\caption{
Dynamics of a pusher-puller microswimmer pair initially located at positions $(x,y)=(0,0)$ and $(4L,0)$.
Swimming trajectories described by the position of the swimmers $\overline{\mathbf{r}}_i$ for a oscillatory spiral state that can further be categorized into (a) avoiding and (c) confronting spiral cases.
The color in the trajectories gradually changes from blue to yellow over time.
Corresponding time evolution of the inter-swimmer distance $\ell$ are presented in panels (b) and~(d).
Here, we set $\mu=0.1$ and $\alpha=2/5$.
The initial orientations for (a,b) and (c,d) are set $(\phi_1,\phi_2)=(2\pi/5,2\pi/5)$ and $(8\pi/5,4\pi/5)$, respectively.
In panels (b) and (d), the insets show the evolution of $\ell$ over a longer period of time.
\label{fig:pusherpuller}
}
\end{figure}

In most of the observed cases, the time period of the resulting diverging spiral state can be approximated by that of an isolated swimmer given by Eq.~\eqref{eq:speed-rotation-rate}, i.e. by neglecting the fluid-mediated hydrodynamic influence of the nearby swimmer.
By using the set of parameters chosen here for the viscosity ratio~$\mu$ and asymmetry parameter~$\alpha$, we obtain $T \left[f_0/(16\pi\eta_{\rm s}L)\right] = 2\pi \alpha \left( 1-\alpha \right)/\mu \approx 15$, which agrees well with the time period observed in Fig.~\ref{fig:pusherpusher}(b).
We note that for some of the initial orientation angles, the time period is found to be somehow smaller than this estimated value.
This generally occurs when the two swimmers are initially aligned along the same direction, a situation for which the influence of the flow field induced by the neighboring swimmer becomes important and thus needs to be accounted for to estimate the time period accurately.

%%%%%%%%%%%%%%%%%%%%%%%%%%%%%%%%
\subsection{Puller-puller swimmer pair}

We first begin with the simplistic situation in which the initial configuration is symmetric such that $\phi_1=\phi_2$.
As shown in Fig.~\ref{fig:pullerpuller}(a), the puller-puller swimmer pair shows in this particular case a \enquote{stable co-orbiting state} in which the binary-swimmer, after a transient evolution, co-rotates in a clockwise direction along circular paths around a common center.
At later times, the inter-swimmer distance converges progressively to a constant value $\ell\approx10L$ given by the orbit diameter; see Fig.~\ref{fig:pullerpuller}(b).
The steady inter-swimmer distance can conveniently be estimated as twice the radius of the circular trajectory obtained for a single swimmer [c.f. Eq.~(\ref{eq:radius})].

For a general initial configuration, the puller-puller swimmer pair exhibits a rather chaotic spiral behavior, which we denote as an \enquote{unstable co-orbiting state}; see Fig.~\ref{fig:pullerpuller}(c) for an exemplary pair dynamics.
We find that the two swimmers first move along spiral trajectories but do not end up trapped in a stable orbit.
In addition, the mutual distance between the two swimmers varies in a non-periodic manner, as presented in Fig.~\ref{fig:pullerpuller}(d).  
During their spiral movement, the two swimmers are constantly attracted toward each other and tend to collide.
However, as the two swimmers get closer to one another, the soft excluded-volume interaction that acts at short inter-particle distances leads to mutual repulsion and subsequent scattering of the two swimmers.

%%%%%%%%%%%%%%%%%%%%%%%%%%%%%%%%
\subsection{Pusher-puller swimmer pair}

For a pair of swimmers with non-identical propulsion mechanisms, we find a period \enquote{oscillating spiral} state, as shown in Fig.~\ref{fig:pusherpuller}(a) and (c) for two exemplary trajectories.
Both swimmers follow spiral trajectories with motion in an anti-clockwise direction for the pusher and in a clockwise direction for the puller.
In addition, the inter-swimmer distance $\ell$ oscillates periodically in time [see Fig.\ref{fig:pusherpuller}(b) and (d)] without diverging on average, as it is found to be the case for the pusher-pusher case discussed above.
Depending on the initial orientation angles, the inter-swimmer distance follows a sine-like evolution in time [Fig.~\ref{fig:pusherpuller}(b)] or features a more complex dynamics during which the swimmer pair comes in close vicinity of each other frequently [Fig.~\ref{fig:pusherpuller}(d)].
Similar to the pusher-pusher case, the period of oscillations obtained in Figs.~\ref{fig:pusherpuller}(c) and (d) can well be approximated by that of an isolated swimmer.

%%%%%%%%%%%%%%%%%%%%%%%%%%%%%%%%
\subsection{Chirality measure}

We quantitatively estimate the chirality of the trajectories in the microswimmer pair dynamics using the measure of chirality $\bar{\zeta}$ introduced in Eq.~(\ref{eq:curvature}). 
For periodic trajectories such as the pusher-pusher and pusher-puller pair dynamics, as shown in Figs.~\ref{fig:pusherpusher} and \ref{fig:pusherpuller}, respectively, we compute $\bar{\zeta}$ by averaging the instantaneous curvature over $10$ periodic iterations.
For the non-periodic case (the puller-puller pair dynamics as in Fig.~\ref{fig:pullerpuller}), we calculate $\bar{\zeta}$ by averaging $\zeta$ from $t\left|f_1\right|/(16\pi\eta_{\rm s}L)=100$ to $200$, a range in which the swimming dynamics is in a steady state or still exhibits a chaotic behavior.
In this basis, the chirality of the pusher-pusher pair with the initial conditions $(\phi_1,\phi_2)=(\pi,\pi)$ is estimated to be $L\bar{\zeta}\approx0.21$, a value comparable to that of a single swimmer $L\bar{\zeta}_0\approx0.21$, suggesting that the pair exhibits trajectories of similar chirality to a single swimmer. 
This similarity can be attributed to the diverging behavior of the pusher-pusher pair and the decaying hydrodynamic interaction between the swimmers over time.
For the puller-puller pair with the initial conditions $(\phi_1,\phi_2)=(6\pi/5,6\pi/5)$, the estimated chirality is $L\bar{\zeta}\approx0.20$, also comparable to that of a single swimmer. 
However, with the parameters $(\phi_1,\phi_2)=(2\pi,9\pi/5)$, the chirality increases to $L\bar{\zeta}\approx0.35$. 
This increased chirality stems from the frequent collisions between the swimmers, which allow circular trajectories with smaller radii.
For the pusher-puller pair, the estimated chirality is $L\bar{\zeta}\approx0.22$ and $L\bar{\zeta}\approx0.26$ for the parameters $(\phi_1,\phi_2)=(2\pi/5,2\pi/5)$ and $(\phi_1,\phi_2)=(8\pi/5,4\pi/5)$ respectively. 
The larger chirality in the latter case is due to the pair dynamics, which bring the swimmers in close together.

%%%%%%%%%%%%%%%%%%%%%%%%%%%%%%%%%%
\section{Summary and Discussion}
%%%%%%%%%%%%%%%%%%%%%%%%%%%%%%%%%%
\label{sec:summary}

In this paper, we have investigated theoretically and by means of computer simulations the low-Reynolds-number hydrodynamics of single and pair of active microswimmers surfing on a 2D compressible odd-viscous fluid layer supported by a 3D fluid underneath.
We have described the solution for the fluid flow induced by a concentrated point-force singularity acting within the fluid layer in terms of the Green's function.
The latter includes a non-reciprocal component resulting from non-vanishing odd viscosity characterizing the 2D layer.
As opposed to the well-studied behavior of an isolated microswimmer in a Newtonian fluid, we have demonstrated that odd viscosity induces finite transverse velocity in addition to rotational flows, leading to swimming along a circular path.
The observed chiral dynamics is a direct consequence of the non-reciprocal component of the Green's function.
Furthermore, we have found that the velocity component parallel to the swimming axis shows a monotonically decreasing dependence on the odd viscosity.
In contrast to that, both the transverse velocity and rotation rate exhibit a non-monotonic behavior.
In the limit of very small odd viscosity relative to the shear viscosity, we have obtained asymptotic analytical expressions of the translational and rotational speeds, in the special case of $\eta_\mathrm{d} = 3\eta_\mathrm{s}$, valid when the overall displacement of the swimmer remains relatively small such that $\kappa r \ll 1$.

In addition, we have demonstrated that two nearby microswimmers show various non-trivial chiral pair dynamics, depending on the initial relative orientation angles and propulsion mechanism (pusher vs.\ puller).
For finite values of odd viscosity, a pair of pushers is found to follow a diverging spiral state.
A pair of pullers either exhibits a stable co-orbiting state, in which the mutual distance between the two pullers reaches a steady value, or a complex chaotic spiral behavior resulting from the interplay of far-field hydrodynamic interactions and near-field steric interactions.
Meanwhile, a pusher-puller pair is found to display an oscillating spiral behavior in which the inter-swimmer distance evolves periodically in time.
Even though we have employed here a rather simple microswimmer model that is composed by two active forces acting on both sides of the body of the swimmer, the observed circular dynamics is expected to manifest itself for variety of other types of self-propelling active agents including squirmers and self-diffusiophoretic Janus colloids.

In the limits of infinitely small swimmer sizes $(L\to0)$ and for the symmetric case $(\alpha=1/2)$, the model used in the present work reduces to the standard force-dipole model that has been used widely to describe the dynamics of enzymatic molecules~\cite{mikhailov2015,illien2017,manikantan2020}. 
In this regime, it would be of interest to study the effects of the odd viscosity on various enzymatic transport phenomena, such as diffusion enhancement~\cite{golestanian2015,illien2017,hosaka2020,hosaka2020_2}, chemotactic or anti-chemotactic behavior~\cite{hosaka2017,agudo2018enhanced,adeleke2019_2}, and synchronization effects~\cite{agudo2021synchronization}.
For instance, the non-reciprocal linear response due to odd viscosity can lead to additional correction terms in the diffusion coefficient of passive tracers in solutions and biological membranes~\cite{mikhailov2015}.

We have not considered here the swimming behavior at large distances ($\kappa r \gg 1$).
In this regime, 3D hydrodynamic effects dominate over 2D effects, resulting in a swimmer dynamics that is governed solely by the viscous properties of the underlying 3D fluid [see Eq.~(\ref{eq:c123large})]~\cite{oppenheimer2010,ramachandran2011}.
Accordingly, the contribution stemming from odd viscosity becomes less significant in predicting the overall dynamics.
However, it is worth mentioning that for a 3D fluid characterized by a 3D odd viscosity~\cite{markovich2021,khain2022,yuan2023stokesian}, the latter will necessarily play a non-negligible role in describing the swimming behavior, even in the limit of large separation distances.
A systematic study along these lines is mandated to investigate both passive and active transport in 3D odd-viscous fluids.

We now provide some estimates of the physical quantities based on the model used in the present work.
By considering a typical microswimmer size $L\sim\SI[parse-numbers=false]{10^{-6}}{\metre}$, propelled through a thrust force $f\sim\SI[parse-numbers=false]{10^{-11}}{\newton}$, and moving in a thin fluid layer of shear viscosity $\eta_{\rm s}\sim\SI[parse-numbers=false]{10^{-6}}{\pascal\second\metre}$~\cite{barentin1999}, the resulting translational and rotational speeds are estimated as $V\sim\SI[parse-numbers=false]{10^{-7}}{\metre\per\second}$ and $\Omega\sim\SI[parse-numbers=false]{10^{-1}}{\per\second}$, respectively
[c.f.~Eq.~(\ref{eq:speed-rotation-rate})].
Based on the model parameters, the predicted radius of the circular trajectory followed by an isolated microswimmer, as given by Eq.~(\ref{eq:radius}), is expected to be of the same order of magnitude as the swimmer size.
In microrheology experiments, the radius of the circular path followed by a swimmer and the rotational frequency at which spinning motion occurs, can potentially serve as a new measurement tool to characterize active transport in odd-viscous systems.
This approach complements the existing protocols for the experimental determination of odd viscosity, such as measurements of the Hall angle~\cite{reichhardt2022}, non-vanishing torque~\cite{ganeshan2017}, or edge current~\cite{soni2019}.
We note that in synthetic systems of rotating colloidal suspensions~\cite{soni2019}, by tuning the externally applied torque, the odd viscosity of the thin fluid layer can be varied.

Experimental evidence has shown that non-reciprocal interactions in chiral algae and embryos~\cite{drescher2009dancing, tan2022odd} or colloidal spinners~\cite{bililign2022motile} lead to self-organization in chiral crystals where rotational motion of the inclusions is an essential prerequisite to form such large assemblies.
We believe that our theoretical predictions can potentially be verified in real space experiments involving biological or artificial microswimmers.
Finally, it would be of great interest to examine the collective behavior of a suspension of many microswimmers in an odd-viscous fluid.
Another interesting aspect that could be addressed is the active transport in systems of curved membrane geometries for which mobility tensors have been derived earlier~\cite{henle2010}.
These are interesting research questions that are worth considering in future investigations.

Direct simulation methods that numerically solve the Navier-Stokes equations are effective for explicitly accounting for the hydrodynamic interaction between swimming bodies. 
The smoothed profile method is one such numerical technique that provides a scheme for coupling continuum fluid mechanics with dispersed moving particles~\cite{yamamoto2021smoothed}. 
Besides directly solving the hydrodynamic equations, computational fluid dynamics methods, such as dissipative particle dynamics~\cite{hoogerbrugge1992simulating}, the lattice Boltzmann method~\cite{ladd2001lattice}, and multi-particle collision dynamics~\cite{lee2004friction}, can also capture hydrodynamic interactions with high accuracy.
However, in this work, our aim was to analyze active swimming behavior using a minimal model for analytical tractability, and therefore we did not employ these numerical techniques.
Furthermore, extensions of existing numerical simulation methods to a system of swimming bodies in a fluid with odd viscosity have rarely been performed~\cite{fruchart2023odd}.
Although such numerical methods are beyond the scope of this work, their application could provide new insights into the effect of odd viscosity on the active dynamics of microswimmers, and as such, they represent an area for future investigation.

%%%%%%%%%%%%%%%%%%%%%%%%%%
\begin{acknowledgments}
%%%%%%%%%%%%%%%%%%%%%%%%%%

We acknowledge support from the Max Planck Center Twente for Complex Fluid Dynamics, the Max Planck School Matter to Life, and the MaxSynBio Consortium, which are funded jointly by the Federal Ministry of Education and Research (BMBF) of Germany and the Max Planck Society.
\end{acknowledgments}

\appendix
\section{Derivation of $C_1$, $C_2$, and $C_3$ when $\eta_{\rm d}\neq3\eta_{\rm s}$}
\label{app:Ci}

Here, we obtain the radial functions $C_1$, $C_2$, and $C_3$ for arbitrary values of $\eta_{\rm s}$ and $\eta_{\rm d}$.
By calculating $G_{ii}$, $G_{ij}\hat{r}_i\hat{r}_j$, and $G_{ij}\epsilon_{ij}$ and performing the inverse Fourier transform of $\widetilde{\mathbf{G}} [\mathbf{k}]$ in Eq.~(\ref{eq:G}), we readily obtain~\cite{hosaka2021}
\begin{subequations} \label{eq:appc1c2c3}
\begin{align}
C_1
&=
\int_0^\infty \,
\frac{{\rm d}k}{2\pi} \,
k \,
\frac{
\eta_{\rm s}(k^2+\kappa^2)J_1(kr)/(kr)
+2\bar{\eta}(k^2+\lambda^2)
[J_0(kr)-J_1(kr)/(kr)]}
{2\eta_{\rm s}\bar{\eta}(k^2+\kappa^2)(k^2+\lambda^2)+\eta_{\rm o}^2k^4},\\
C_2&=
\int_0^\infty 
\frac{{\rm d}k}{2\pi} \,
k \,
\frac{ 
[\eta_{\rm s}(k^2+\kappa^2)
-2\bar{\eta}(k^2+\lambda^2)
]
[J_0(kr)-2J_1(kr)/(kr)]
}
{2\eta_{\rm s}\bar{\eta}(k^2+\kappa^2)(k^2+\lambda^2)+\eta_{\rm o}^2k^4},
\\
 C_3 &=
-\eta_{\rm o}
\int_0^\infty 
\frac{{\rm d}k}{2\pi}
\frac{k^3J_0(kr)}{2\eta_{\rm s}\bar{\eta}(k^2+\kappa^2)(k^2+\lambda^2)+\eta_{\rm o}^2k^4}.
\end{align}
\end{subequations}
When $\eta_{\rm d}=3\eta_{\rm s}$ or equivalently $\kappa=\lambda$, we obtain Eqs.~\eqref{eq:c}.

\section{Derivation of Eqs.~(\ref{eq:twoV}) and (\ref{eq:twoOmega})}
\label{app:vomega}

We derive the translational velocities of the two hydrodynamically interacting swimmers in Eq.~(\ref{eq:twoV}) and the angular velocities in Eq.~(\ref{eq:twoOmega}).
Substituting Eq.~(\ref{eq:inducedvi}) into Eq.~(\ref{eq:totalv}) with the use of Eq.~(\ref{Eq:VeloField}) and evaluating the resulting velocity at the position $\overline{\mathbf{r}}_i$, we obtain the velocity of the swimmer $i$ subject to the flow induced by the swimmer $j$ as
\begin{align}
\mathbf{V}_i &=  
\frac{5\beta}{16\pi\eta_{\rm s}}
\left(\mathbf{I}-\frac{2}{5} \, \mu \boldsymbol{\epsilon} \right)\cdot
\mathbf{f}_i
+
\frac{5}{16\pi\eta_{\rm s}}
\left[
\ln\frac{|\mathbf{r}_{ij}+(1-\alpha)L\mathbf{t}_j|}{|\mathbf{r}_{ij}-\alpha L\mathbf{t}_j|}
\left(\mathbf{I}-\frac{2}{5}\mu \boldsymbol{\epsilon} \right)
+\right.\nonumber\\
&+\left. \, \frac{3}{5} 
\left( 
\frac{(\mathbf{r}_{ij}-\alpha L\mathbf{t}_j)(\mathbf{r}_{ij}-\alpha L\mathbf{t}_j)}{|\mathbf{r}_{ij}-\alpha L\mathbf{t}_j|^2}
-
\frac{[\mathbf{r}_{ij}+(1-\alpha)L\mathbf{t}_j][\mathbf{r}_{ij}+(1-\alpha)L\mathbf{t}_j]}{|\mathbf{r}_{ij}+(1-\alpha)L\mathbf{t}_j|^2} \right)
\right]
\cdot\mathbf{f}_j,
\label{eq:appvi}
\end{align}
where the first term represents the self-interaction contribution from Eq.~(\ref{eq:speed-rotation-rate}) and the second term represents the pair-interaction contribution from Eq.~(\ref{Eq:VeloField}).
Defining $\mathbf{K}_{ij}^+ = \mathbf{r}_{ij} - \alpha L \hat{\mathbf{t}}_j$ and $\mathbf{K}_{ij}^- = \mathbf{r}_{ij} + \left(1-\alpha\right) L \hat{\mathbf{t}}_j$ in Eq.~(\ref{eq:appvi}) and deriving the expressions for $\mathbf{V}_1$ and $\mathbf{V}_2$, we obtain Eq.~(\ref{eq:twoV}).

To derive the angular velocity of a swimmer interacting with another swimmer, we first calculate the gradient of the velocity field induced by the swimmer $i$ in Eq.~(\ref{Eq:VeloField})
\begin{align}
    \partial_\alpha v_{i,\beta}(\mathbf{r})
    &=
    \frac{5}{16\pi\eta_{\rm s}}
    \left[
    \left(
    \frac{R_{-,\alpha}}{R_-^2}
    -
    \frac{R_{+,\alpha}}{R_+^2}
    \right)
    \left(
    \delta_{\beta\gamma}
    -
    \frac{2}{5}
    \mu
    \epsilon_{\beta\gamma}
    \right)
    +
    \frac{3}{5}
    \left(
    \frac{R_{+,\beta}}{R_+^2}
    -
    \frac{R_{-,\beta}}{R_-^2}
    \right)
    \delta_{\alpha\gamma}
    \right.\nonumber\\
    &
    \left.+
    \frac{3}{5}
    \left(
    \frac{R_{+,\gamma}}{R_+^2}
    -
    \frac{R_{-,\gamma}}{R_-^2}
    \right)
    \delta_{\alpha\beta}
    -
    \frac{6}{5}
    \left(
    \frac{R_{+,\alpha}R_{+,\beta}R_{+,\gamma}}{R_+^4}
    -
    \frac{R_{-,\alpha}R_{-,\beta}R_{-,\gamma}}{R_-^4}
    \right)
    \right]f_{i,\gamma},
\end{align}
where we have assumed summation over repeated Greek indices.
The $z$-component of the half of the vorticity induced by the swimmer $i$ can be written as
\begin{align}
    \frac{1}{2}
    \epsilon_{\alpha\beta}
    \partial_\alpha v_{i,\beta}(\mathbf{r})
    =
    \frac{1}{16\pi\eta_{\rm s}}
    \left(
    \frac{R_{-,\alpha}}{R_-^2}
    -
    \frac{R_{+,\alpha}}{R_+^2}
    \right)
    \left(
    \mu
    \delta_{\alpha\beta}
    +
    4
    \epsilon_{\alpha\beta}
    \right)
    f_{i,\beta}.
    \label{eq:appvorticity}
\end{align}
Then, the $z$-component of the angular velocity of the simmer $1$ can be expressed with the sum of Eq.~(\ref{eq:appvorticity}) both for the swimmer $1$ and $2$, which is evaluated at $\mathbf{r}=\overline{\mathbf{r}}_1$
\begin{align}
    \Omega_1
    =
    \left.
    \frac{1}{2}
    \nabla\times
    \left[
    \mathbf{v}_{1}(\mathbf{r})
    +
    \mathbf{v}_{2}(\mathbf{r})
    \right]_z
    \right|_{\mathbf{r}=\overline{\mathbf{r}}_{1}}
    =
    \frac{1}{16\pi\eta_{\rm s}}
    \left[
    \frac{\mu \hat{\mathbf{t}}_1\cdot\mathbf{f}_1}{\alpha(1-\alpha)L}
    +
    \left(
    \frac{\mathbf{K}_{12}^-}{\left|\mathbf{K}_{12}^-\right|^2}
    -
    \frac{\mathbf{K}_{12}^+}{\left|\mathbf{K}_{12}^+\right|^2}
    \right)
    \cdot
    \left(
    \mu
    \mathbf{I}
    +
    4
    \boldsymbol{\epsilon}
    \right)
    \cdot
    \mathbf{f}_2
    \right].
\end{align}
Noting $\mathbf{f}_1=\hat{\mathbf{t}}_1f_1$ and deriving the similar expression for $\Omega_2$, we obtain Eq.~(\ref{eq:twoOmega}).

% \bibliography{myref} % Entries are in the refs.bib file

% \begin{comment}
%aipnum4-2.bst 2019-01-14 (MD) hand-edited version of apsrev4-1.bst

% \end{comment}

\end{document}